\documentclass[10pt]{article}

\usepackage[utf8]{inputenc} 
\usepackage{amsmath}
\usepackage{amssymb}
\usepackage{mathrsfs}
\usepackage{bm}
\usepackage{color}
\usepackage{caption}

\usepackage[margin=1in,footskip=0.25in]{geometry}
\usepackage{graphicx}
\DeclareGraphicsExtensions{.pdf,.jpg}

\usepackage{caption}
\usepackage{subcaption}

\newcommand{\ket}[1]{| #1 \rangle}
\newcommand{\bra}[1]{\langle #1 |}

\newcommand{\tr}{\text{Tr}}
\newcommand{\q}[1]{\vec{#1}\cdot\vec{\sigma}}

\begin{document}

\title{Boltzmann-Gibbs states in topological quantum walks and associated many-body systems:
\\
fidelity and Uhlmann parallel transport analysis of phase transitions}
\maketitle

\author{Bruno Mera$^{1,2,3}$, C. Vlachou$^{4,5}$, N. Paunkovi\'c$^{4,5}$ 
and V\'{\i}tor R. Vieira$^{1,2}$}
\begin{center}
$^1$ CeFEMA, Instituto Superior
T\'ecnico, Universidade de Lisboa, Av. Rovisco Pais, 1049-001 Lisboa, Portugal\\
$^2$  Departamento de F\'{\i}sica,  Universidade de Lisboa, Av. Rovisco Pais, 1049-001 Lisboa, Portugal\\
 $^3$  Physics of Information and Quantum Technologies Group, Instituto de Telecomunica\c{c}\~oes, 1049-001 Lisbon, Portugal\\
$^4$ SQIG-Security and Quantum Information Group, Instituto de Telecomunica\c{c}\~oes, 1049-001 Lisbon, Portugal\\
$^5$ Departamento de Matemática, Instituto Superior Técnico, Universidade de Lisboa, Av. Rovisco Pais, 1049-001 Lisboa, Portugal
\end{center}
\begin{abstract}
We perform the fidelity analysis for Boltzmann-Gibbs-like states in order to investigate whether the topological order of 1D fermionic systems at zero temperature is maintained at finite temperatures. We use quantum walk protocols that are known to simulate topological phases and the respective quantum phase transitions for chiral symmetric Hamiltonians.  Using the standard approaches of the fidelity analysis and the study of edge states, we conclude that no thermal-like phase transitions occur as temperature increases, i.e., the topological behaviour is washed out gradually. We also show that the behaviour of the Uhlmann geometric factor associated to the considered fidelity exhibits the same behaviour as the latter, thus confirming the results obtained using the previously established approaches.
\end{abstract}
\maketitle
\section{Introduction}
\label{sec:introduction}
Since the seminal paper of Haldane~\cite{hal:88}, where the anomalous Hall insulator was discovered, there has been an intense investigation on topological phases of matter. To characterise topological phases, invariants such as Chern numbers~\cite{tknn:82} and the Berry geometric phase~\cite{ber:84}, as well as non-local string parameters~\cite{zen:che:zho:wen:15}, were used as signatures of topological orders (cf.~\cite{ando:13}). Regarding zero-temperature quantum phase transitions, the critical behaviour of systems featuring topological order is accompanied by the existence of edge states at the boundary between two distinct topological phases~\cite{x:g:wen:91,ryu:hat:02}. Moving to open quantum systems, that are described by mixed, rather than pure, states, it is natural to ask if the topological order is still present, and investigate which are the appropriate quantities (``topological order parameters'') to describe possible phase transitions. Various types of the mixed-state generalisations of geometric phases~\cite{uhl:89,sjo:15} and closely related quantities~\cite{viy:riv:del:14,viy:riv:del:2d:14,viy:riv:del:15,zho:aro:14}, as well as the Chern values~\cite{riv:viy:del:13}, were used to infer topological phase transitions of open systems (thermal and/or non-equilibrium). 
	
	In addition to the standard local symmetry breaking and the aforementioned global topological order parameters, several information-theoretic quantities were used to study phase transitions, such as entanglement measures~\cite{ham:ion:zan:05,ham:zha:haa:lid:08,hal:ham:12} and the fidelity~\cite{aba:ham:zan:08,zan:pau:06,zha:zho:09, oli:sac:14, pau:sac:nog:vie:dug}. Being the measure of the distinguishability between two quantum states, the fidelity was used in numerous studies of both zero- and finite-temperature symmetry-breaking phase transitions, as well as of Kosterlitz-Thouless transitions~\cite{gu:kwo:nin:wen:lin:08,maz:ham:12}. In~\cite{zan:gio:coz:07} the authors analysed the intimate connection between the pure-state fidelity and the Berry phase, showing that the fidelity-induced Riemannian metric and the Berry curvature are the real and imaginary part, respectively, of the so-called quantum geometric tensor. Regarding mixed thermal states, the fidelity and the Uhlmann factor were studied for the case of the BCS superconductivity~\cite{pau:vie:08} and subsequently, in addition to BCS, for topological insulators and superconductors~\cite{mer:vla:pau:vie:17}.
	
	Recently, Kitagawa {\em et al.}~\cite{kit:rud:ber:dem:10}, showed that discrete-time quantum walks (QW) can realise topological phases in 1D and 2D for all the symmetry classes~\cite{sch:ryu:fur:lud:08,kit:09} of free-fermion systems. In particular, they provide the QW protocols that simulate representatives of all topological phases, featured by the presence of robust symmetry-protected edge states (see also~\cite{kit:12}). In general, QW realisations are particularly useful, because, in addition to the simplicity of their mathematical description, the parameters that define them can be easily controlled in the lab. The aforementioned topological quantum walks have been experimentally realised as periodically driven systems~\cite{kit_exp:12} and there are several experimental proposals for measuring topological invariants employing this approach~\cite{rak:asb:alb:16,gro:bra:alt:mes:asb:16,mug:cel:mas:asb:lew:lob:16}.

We study topological features at finite temperatures for Boltzmann-Gibbs (BG) thermal-like states of single-particle 1D discrete-time topological QW and the corresponding many-body systems. The topological QW protocols we use simulate representatives of the chiral symmetric classes BDI and AIII of 1D topological insulators. We analyse the behaviour of the fidelity, the associated Uhlmann parallel transport and the edge states, and conclude that the effective temperature only smears out the topological features exhibited at zero temperature, without causing  temperature-driven phase transitions. 	
		
This paper is organised as follows: in the next section, we describe the main topological features of QWs and their origin, and present the respective protocols. For a detailed and complete analysis see~\cite{kit:rud:ber:dem:10,kit:12}. In Section~\ref{sec:density_operators} we present the BG states used: the single-particle QW states and their many-body counterparts. Furthermore, we clarify the relationship between them and explain the motivation for their use in different physical scenarios. In Section~\ref{sec:results}, we present our results on the fidelity, the Uhlmann parallel transport and the edge states at finite temperatures and discuss the possibility of temperature-driven phase transitions. Finally, we summarise and discuss the results and point out possible directions of future work.  

\section{Topological quantum walks}
\label{sec:top_qw}
We consider the {\em split-step} 1D discrete-time QW introduced in~\cite{kit:rud:ber:dem:10}. A single step is given by the unitary 
\begin{align}
U(\theta_1,\theta_2)=T_{1}R(\theta_2)T_{0}R(\theta_1),
\end{align}
where the shift operators are 
\begin{align}
T_{c}=\sum_{x}\ket{x+(-1)^c}\bra{x}\otimes\ket{c}\bra{c}+\ket{x}\bra{x}\otimes\ket{1\oplus c}\bra{1\oplus c},
\end{align}
with $c\in\{ 0,1\}$ and $\oplus$ being addition modulo 2, and the coin operators are $R(\theta) = e^{i\frac{\theta}{2}\vec\alpha \cdot \vec\sigma}$, where $\vec\alpha$ is a unit 3D vector and $\vec\sigma =(\sigma_x , \sigma_y , \sigma_z)$ is the Pauli vector. 

As mentioned in the Introduction, topological QWs can be realised by means of periodically driven systems given by periodic time-dependent Hamiltonians  $H(t+\delta t)=H(t)$, where $\delta t$ represents the time of a single step. The evolution operator for one period of the driving, $[0,\delta t]$, called the Floquet operator, is given by
\begin{align}
\label{eq:floquet}
U(\delta t)=\mathcal{T}e^{-i\int_{0}^{\delta t}H(t)dt},
\end{align}
where $\mathcal{T}$ is the time ordering operator. Using homotopy theory, in~\cite{kit:ber:rud:dem:10} the authors propose a classification of periodically driven systems, according to the topological properties of their Floquet operators. They consider the Floquet operator in terms of a local effective Hamiltonian, given by $U(\delta t)=e^{-iH_\text{eff}(\delta t)}$. They show that if $U(\delta t)$ is trivial under all homotopy groups, then the associated $H_\text{eff}$ can exhibit non-trivial topological behaviour. The triviality of $U(\delta t)$ under the homotopy groups implies the existence of a gap in the spectrum of $H_\text{eff}$ and if moreover $H_\text{eff}$ has certain symmetries, the system supports the topological phases present in static topological insulators and superconductors, classified according to its dimension and the presence of these symmetries~\cite{sch:ryu:fur:lud:08,kit:09}. 

The unitary operator that describes one step of the evolution of the split-step QW, as given in (1), is trivial under all homotopy groups, therefore we can define a local effective Hamiltonian 
\begin{align}
H_{\text{eff}}(\theta_1,\theta_2) \equiv -i\delta t^{-1}\log U(\theta_1,\theta_2), 
\end{align}
whose quasienergy spectrum has a gap~\cite{kit:ber:rud:dem:10}.
To fix the branch of the logarithm, we choose for energy spectrum the first Brillouin zone, as in~\cite{asb:12,asb:obu:13}, obtaining the single-particle $H_{\text{eff}}$ consistent with realistic many-body counterparts discussed in the next section. This way, the QW ``provides a stroboscopic simulation of the evolution generated by $H_{\text{eff}}$ at the discrete times $N\delta t$''~\cite{kit:rud:ber:dem:10}. In other words, the evolution of the QW is performed in discrete time steps which last $\delta{t}$ units of time each. 

Due to translational invariance, we can write: 
\begin{align}
H_{\text{eff}}(\theta_1,\theta_2) =\sum_k [E_k (\theta_1,\theta_2)\vec{n}_k (\theta_1,\theta_2)\cdot\vec{\sigma}]\otimes\ket{k}\bra{k},
\end{align}
where $\mathcal {B} $ is the first Brillouin zone, $E_k \geq 0$ and $\vec{n}_k$ are 3D unit vectors (for simplicity, we omit the $\theta$-dependences). The eigenstates are given by
 \begin{align}
H_k \ket{\pm\vec{n}_k} = \pm E_k\ket{\pm\vec{n}_k}, \text{ with } H_k=E_k\vec{n}_k\cdot\vec{\sigma}, 	
\label{Eq: Eigensystem of H_k}
 \end{align}
forming the two energy bands $\{\pm E_k, \ k\in \mathcal{B}\}$. This type of Hamiltonians feature topological phases in the presence of time-reversal, particle-hole and/or chiral/sublattice symmetry~\cite{kit:rud:ber:dem:10,kit:ber:rud:dem:10}. The symmetry classes we study in this paper
 (BDI and AIII) are chiral. This implies that $\vec{n}_k$, which defines the quantisation axis for each quasimomentum $k$, is restricted to lie on a great circle of the Bloch sphere, determined by $\vec\alpha$ and $\theta_1$. The number of times that $\vec{n}_k$ winds around the origin, as $k$ ranges within the first Brillouin zone $\mathcal B$, is the {\em winding number} of the map between the two circles. This is exactly what gives rise to the topological features of the QW. The other symmetries are  then determined by the choice of $\vec\alpha$ and/or by doubling the coin degree of freedom~\cite{kit:rud:ber:dem:10,kit:12}. For each symmetry class, we fix the value of $\theta_1$ ($\theta_1=-\pi/2$ for  BDI and $\theta_1=\pi/2$ for AIII) and vary $\theta_2$~\cite{kit:rud:ber:dem:10}, so from now on, we consider $H_{\text{eff},\theta_1}(\theta) \equiv H_{\text{eff}}(\theta_1,\theta)$. By additionally varying $\theta$ along the line of the walk, it is possible to create a domain wall that separates two different phases, where the aforementioned edge states lie (see Appendix 1, for the details).

\section{Boltzmann-Gibbs density operators}
\label{sec:density_operators}
As described in the previous section, QWs can simulate topological phases of free-fermion systems at zero temperature. We wish to investigate if they can also be used to infer topological behaviour of many-body free-fermion systems at finite temperature also. To do that, we study two types of BG-like states for many-body systems, as well as their single-particle counterparts, for which we consider the effective Hamiltonian of the QW. Since topological QWs can be realised as periodically driven systems, it is natural to ask if such states can be considered in this context, since the energy is not conserved and the quasienergies (defined modulo $2\pi$) which are conserved, have no natural ordering. It has been shown that these states, called Floquet-Gibbs states, can emerge under certain conditions~\cite{shi:mor:miy:15,shi:thi:mor:han:may:16} (see also the justification for the existence of a ``quasienergy Brillouin zone'' in the previous section). Following the discussion from the previous section, regardless of the realisations using periodically driven systems, the QWs that we consider can also be achieved using time-independent effective Hamiltonians (which were subject of a number of theoretical studies \cite{kit:rud:ber:dem:10,asb:12,asb:obu:13}, and also experimentally realised~\cite{kit_exp:12,bar_exp:16}), for which the BG states are well defined.

Let us consider a collection of fermion creation and annihilation operators $\{\psi_{k\sigma},\psi^{\dagger}_{k\sigma}: k \in \mathcal B,\ \sigma\in\{\uparrow,\downarrow\}\}$ and form the spinors $\Psi_k=(\psi_{k\uparrow},\psi_{k\downarrow})^T$.
The first state that we consider is the canonical ensemble given by,
\begin{align}
\label{varrho_0}
\varrho^{(0)}=\frac{e^{-\beta \mathcal{H}}}{\mathcal{Z}^{(0)}} = \frac{1}{\mathcal{Z}^{(0)}}\prod_{k\in \mathcal{B}} \exp(-\beta \Psi_k^{\dagger} H_k\Psi_k),	
\end{align}
where $\mathcal{H}$ is given by the sum over the momenta of quadratic Hamiltonians $\mathcal{H} =\sum_k \Psi^{\dagger}_k H_k \Psi_k$, $H_k$ is given by Eq.~\eqref{Eq: Eigensystem of H_k}, and $\mathcal{Z}^{(0)} = \tr(e^{-\beta\mathcal{H}})$ is the corresponding partition function (note that $\mathcal{H}$ conserves the particle number and its action on the whole Hilbert space is determined by the action on the single-particle sector). This state maximises the von Neumann entropy, subject to the constraint $\langle\mathcal{H}\rangle=\text{const}$.

The second one appears when one maximises the von Neumann entropy, subject to two constraints: the above mentioned energy constraint, as well as the constraints on the average number of particles $\langle n_k\rangle=\langle\Psi_k^{\dagger} \Psi_k\rangle$ which are constant in time, but in general different for each $k$. The state is of the form: 
\begin{align}
\label{varrho_1}
\varrho^{(1)}=\frac{e^{-\beta \mathcal{O}}}{\mathcal{Z}^{(1)}} = \frac{1}{\mathcal{Z}^{(1)}} \prod_{k\in \mathcal{B}} \exp[-\beta (\Psi_k^{\dagger} H_k\Psi_k-\mu_k\Psi_k^{\dagger}\Psi_k)],	
\end{align}
where $\mathcal{O}=\sum_k \mathcal{O}_k=\sum_k\mathcal{H}_k-\mu_k n_k,$ with $\mu_k=-(1/\beta)\log(Z_k)$ being a momentum dependent chemical potential (which, in this case, coincides with the Helmholtz free energy associated to momentum $k$), and $\mathcal{Z}^{(1)} = \tr(e^{-\beta\mathcal{O}})$ is the corresponding partition function. For details concerning the derivation of these states or, more generally, states maximising the von Neumann entropy subject to constraints, cf. the appendix of~\cite{vie:10}. In the field of quantum integrable systems, the state $\varrho^{(1)}$ is known as the ``generalised Gibbs ensemble'', which was introduced in~\cite{rig:dun:yur:ols:07,rig:mur:ols:06} (for a review, see~\cite{vid:rig:16}).

Since the previous zero-temperature studies (both theoretical and experimental) of symmetry-protected topological orders were done in terms of single-particle quantum walks, we present the corresponding single-particle projected states $\rho^{(0)}$ and $\rho^{(1)}$ of the many-body counterparts~\eqref{varrho_0} and~\eqref{varrho_1}, respectively. 

The first is the standard thermal state, resulting from the effective Hamiltonian of the QW:
\begin{align}
\label{rho_0}
\rho^{(0)} &= \frac{e^{-\beta H_{\text{eff}}}}{Z}= \frac{1}{Z}\sum_{k \in \mathcal B} e^{-\beta H_k}\otimes \ket{k}\bra{k},
\end{align}
where $Z=\tr e^{-\beta H_{\text{eff}}}$, while the second is:
\begin{align}
\label{rho_1}
\rho^{(1)} = \frac{1}{\Omega}\sum_{k \in \mathcal B} \frac{e^{-\beta H_k}}{Z_k}\otimes \ket{k}\bra{k} = \frac{1}{\Omega}\sum_{k \in \mathcal B} \rho_k\otimes \ket{k}\bra{k},
\end{align}
where $Z_k=\tr e^{-\beta H_k}$ and $\Omega=\sum_k 1$ is the $k$-space volume. Note that by tracing out the momenta, the state $\rho^{(0)}$ remains to be of the BG form (it is ``globally'', with respect to $k$, thermal-like), while $\rho^{(1)}$ is not -- only by measuring the momenta, the state collapses to a ``local'' BG form $\rho_k$. Notice also that $Z=\sum_k Z_k$. The difference between $\rho^{(0)}$ and $\rho^{(1)}$ becomes even more clear by looking at their asymptotic behaviours. Namely, when $\beta\rightarrow +\infty$
\begin{align}
\rho^{(0)}\rightarrow \frac{1}{|M|}\sum_{k\in M} \ket{-\vec{n}_k}\bra{-\vec{n}_k}\otimes \ket{k}\bra{k},
\label{eq:rho0_lim}
\end{align}
where $M=\{k_{*}\in \mathcal B : E(k_*) = \max_{k\in \mathcal B} E(k)\}$ is the set of momenta minimising the lower band dispersion and $\ket{-\vec{n}_k}$ is defined in Eq.~\eqref{Eq: Eigensystem of H_k}, while
\begin{align}
\rho^{(1)}\rightarrow \frac{1}{\Omega}\sum_{k \in \mathcal B} \ket{-\vec{n}_k}\bra{-\vec{n}_k}\otimes\ket{k}\bra{k},	
\label{eq:rho1_lim}
\end{align}
is a statistical mixture of the {\em entire} lower band of the Hamiltonian.

Note that there exists a bijection between single-particle and quadratic many-body Hamiltonians, given by $ H_k\leftrightarrow \mathcal{H}_k = \Psi^{\dagger}_k H_k \Psi_k $,
where the left arrow represents the aforementioned projection onto the single-particle sector, thus inducing the corresponding bijections between the BG states $\rho^{(0)} \leftrightarrow \varrho^{(0)}$ and $\rho^{(1)} \leftrightarrow \varrho^{(1)}$.

\section{The results: fidelity, Uhlmann parallel transport and the edge states}
\label{sec:results}

As mentioned in the Introduction, the fidelity is a measure of the distinguishability between two quantum states $\rho$ and $\rho'$, and is given as $F(\rho,\rho')\equiv \tr\sqrt{\sqrt{\rho}\rho'\sqrt{\rho}}.$ By definition $F(\rho,\rho')\geq 0,$ with the equality holding in the case of two completely distinguishable states, and $F(\rho,\rho') = 1$ if and only if $\rho = \rho',$ which is the maximum value fidelity can take in the case of two completely indistinguishable states. When a system undergoes a phase transition, the density matrix describing it changes drastically, resulting in a sudden drop of the fidelity. This is the main motivation for using fidelity to detect phase transitions. Furthermore, the fidelity is closely related to the Uhlmann connection via the Bures metric over the parameter space (for details, see~\cite{mer:vla:pau:vie:17}). In particular, the ``Uhlmann factor'' $U$, given by $\sqrt{\rho}\sqrt{\rho^\prime} = |\sqrt{\rho}\sqrt{\rho^\prime}|U$, quantifies the difference between the eigenbases of the two states $\rho$ and $\rho^\prime$, and thus the distinguishability between them~\cite{pau:vie:08}. To evaluate the departure of $U$ from identity we consider the quantity
\begin{align}
\Delta(\rho,\rho')=F(\rho,\rho')-\tr(\sqrt{\rho}\sqrt{\rho'}) = \tr(|\sqrt{\rho}\sqrt{\rho^\prime}|(I-U)),
\end{align}
which we also use for the detection of phase transitions.

\begin{figure}[h!]
\center
\begin{minipage}{1.22\textwidth}
\begin{flushleft}
\includegraphics[width=0.20\textwidth,height=0.15\textwidth]{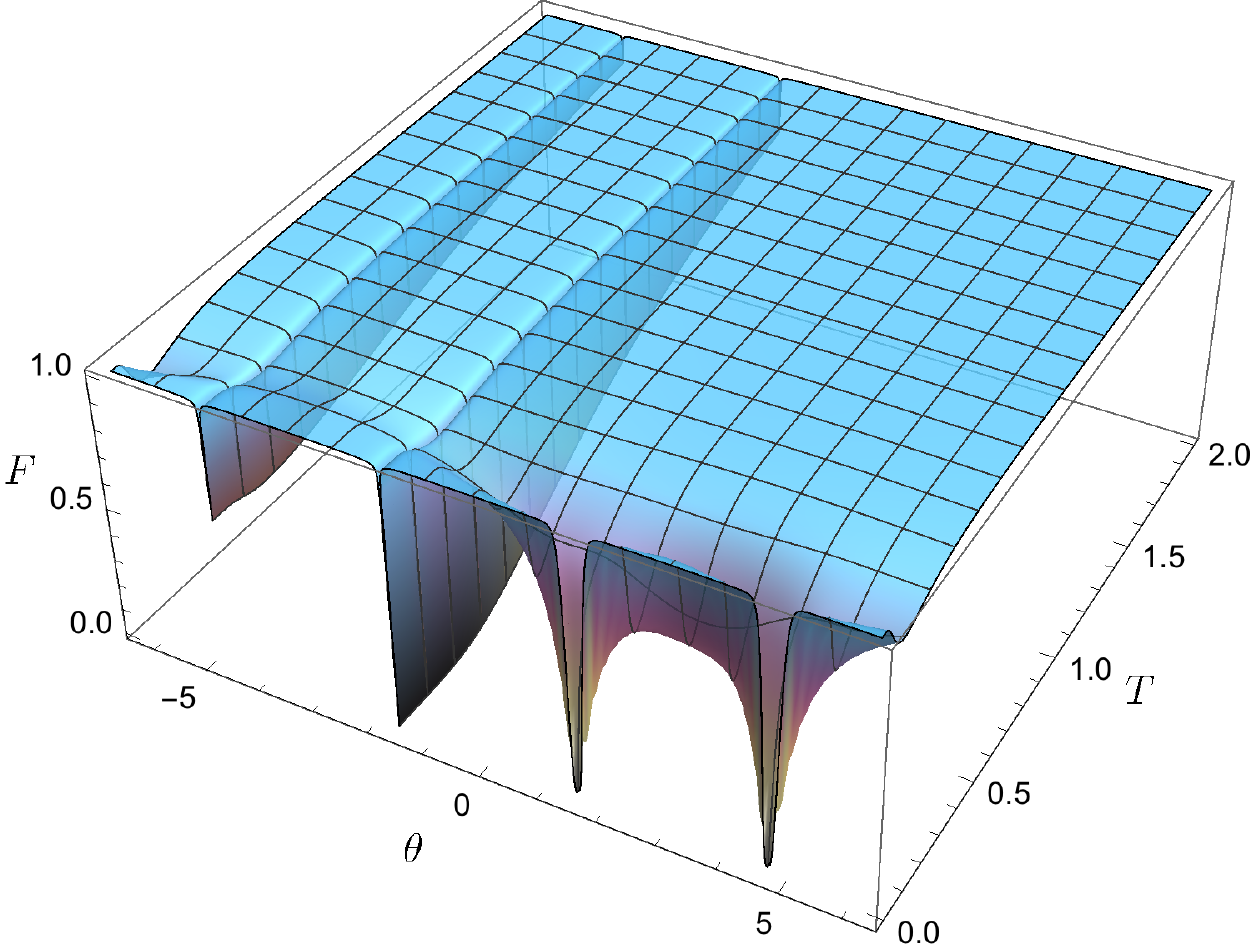}
\includegraphics[width=0.20\textwidth,height=0.15\textwidth]{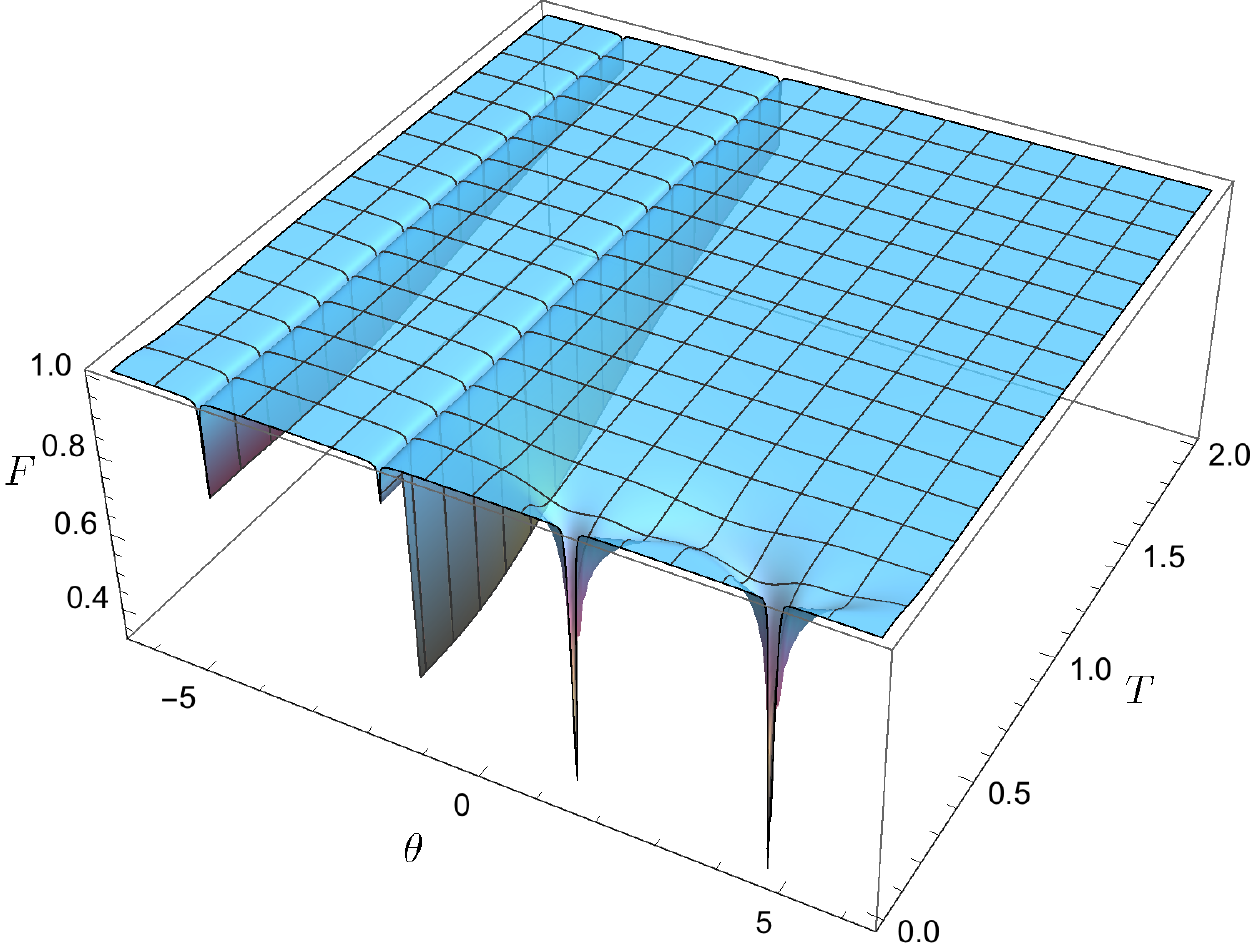}
\includegraphics[width=0.20\textwidth,height=0.15\textwidth]{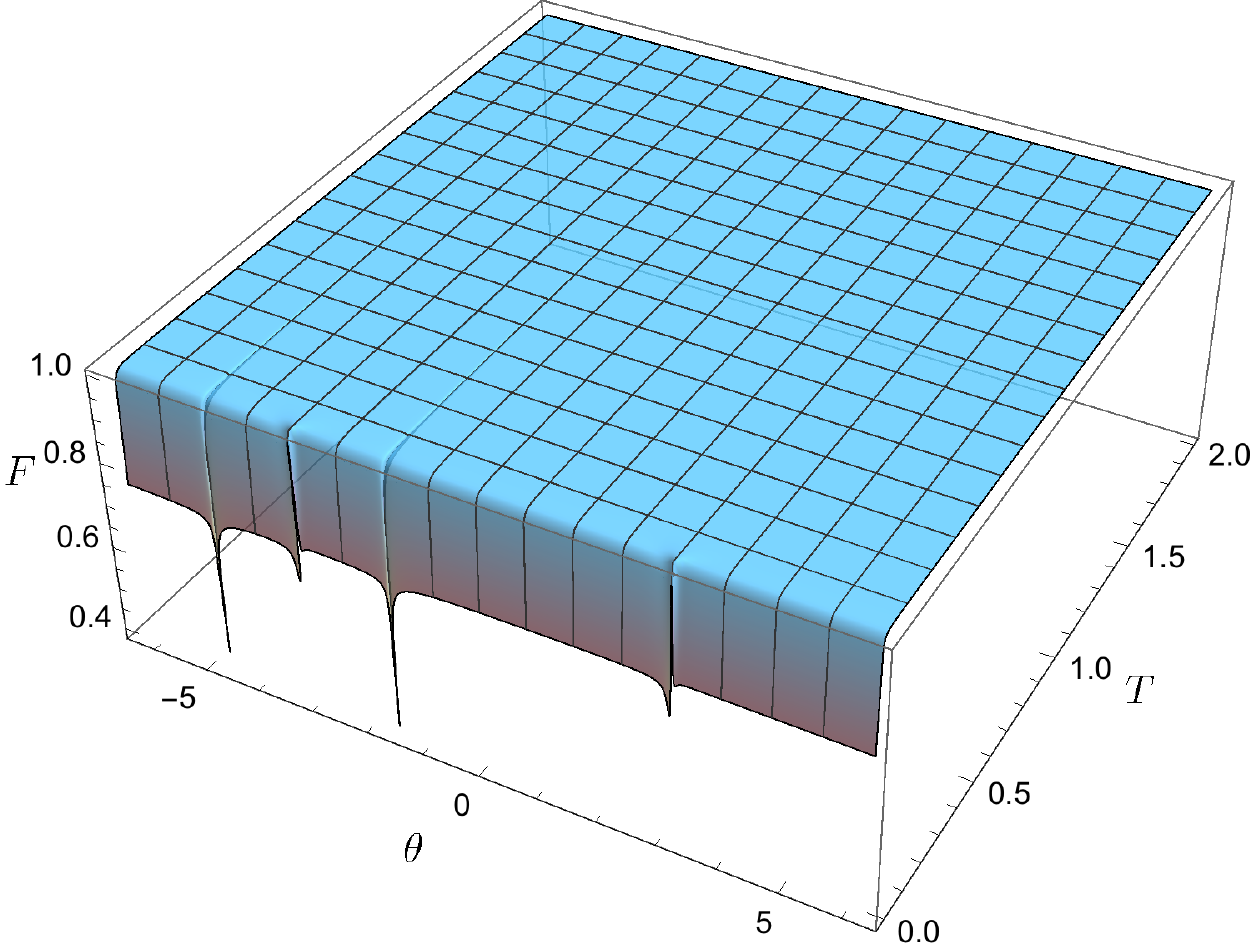}
\includegraphics[width=0.20\textwidth,height=0.15\textwidth]{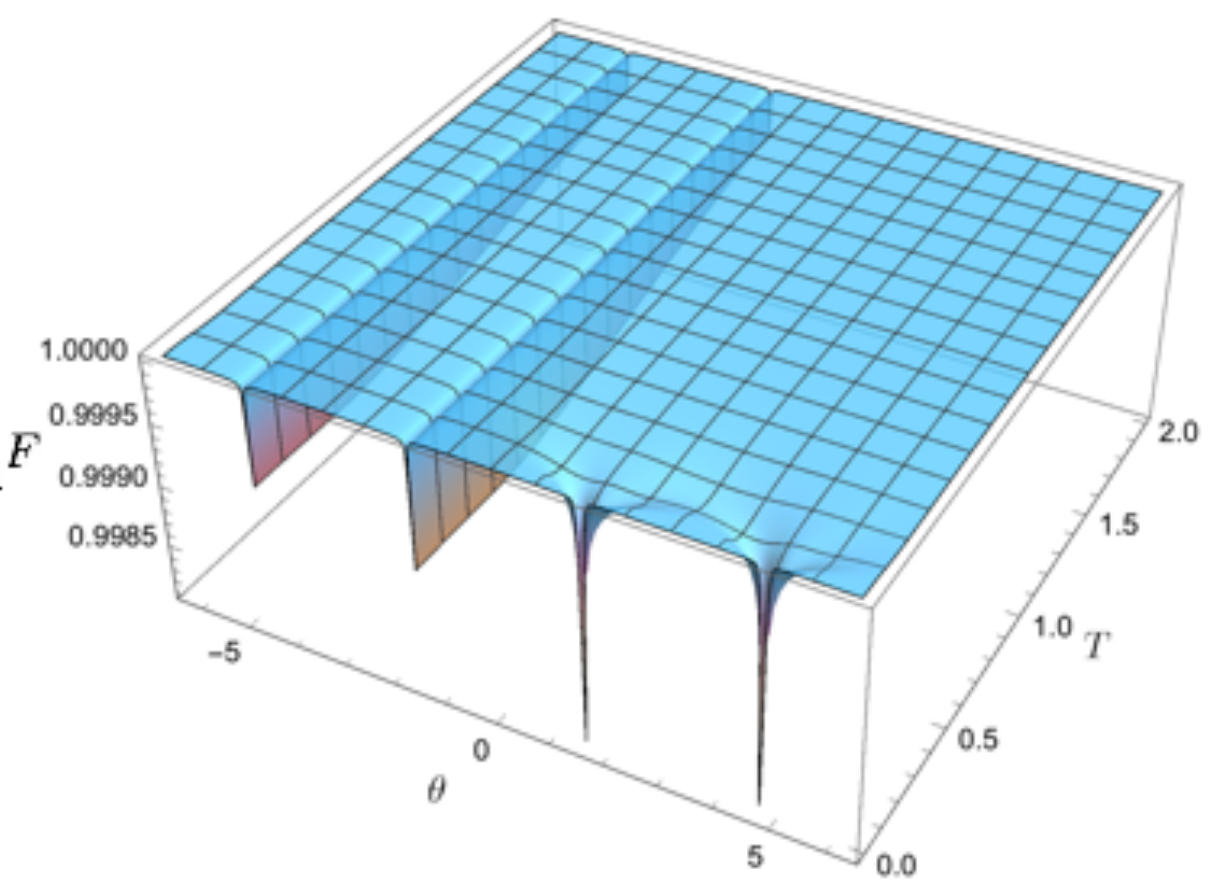}
\end{flushleft}
\end{minipage}
\begin{minipage}{1.22\textwidth}
\begin{flushleft}
\includegraphics[width=0.20\textwidth,height=0.15\textwidth]{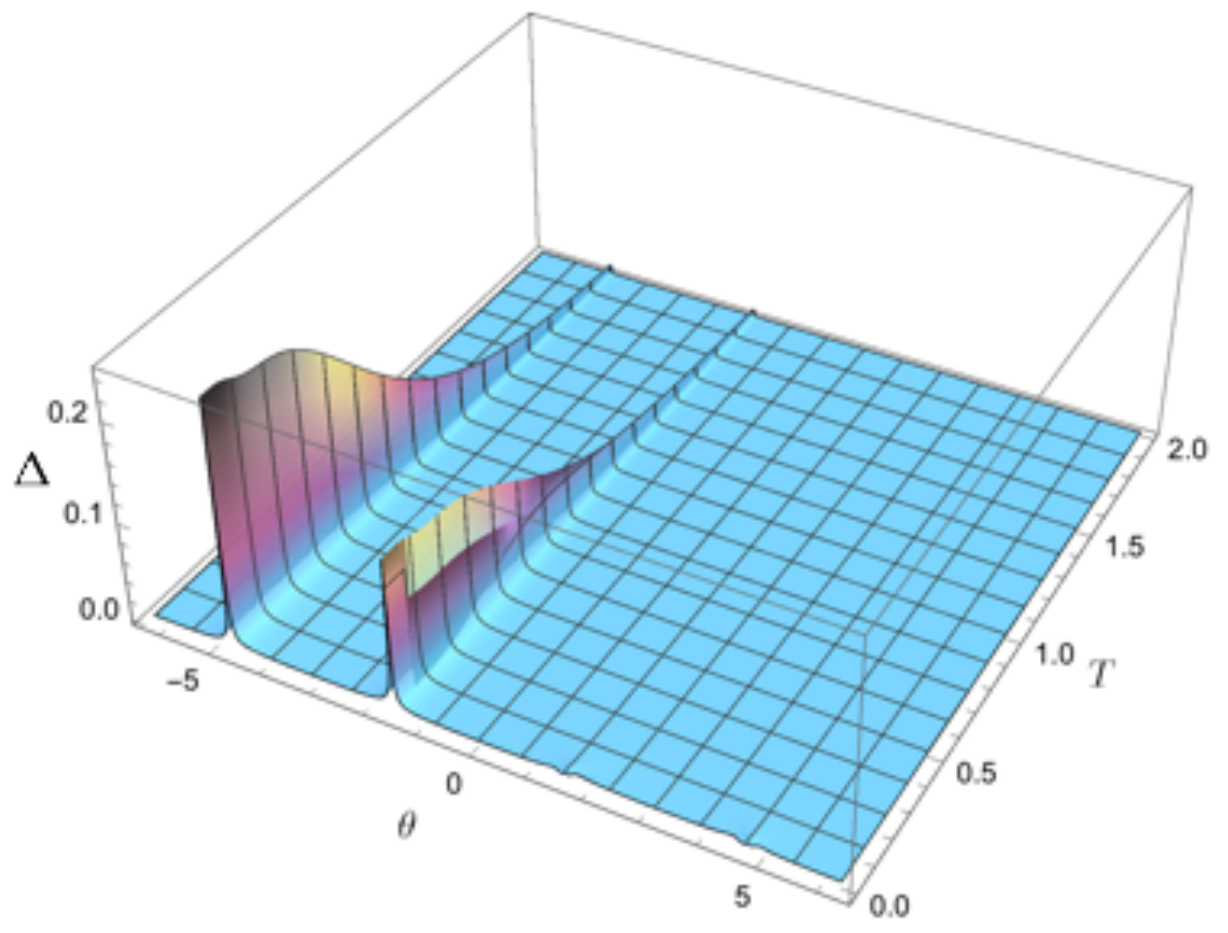}
\includegraphics[width=0.20\textwidth,height=0.15\textwidth]{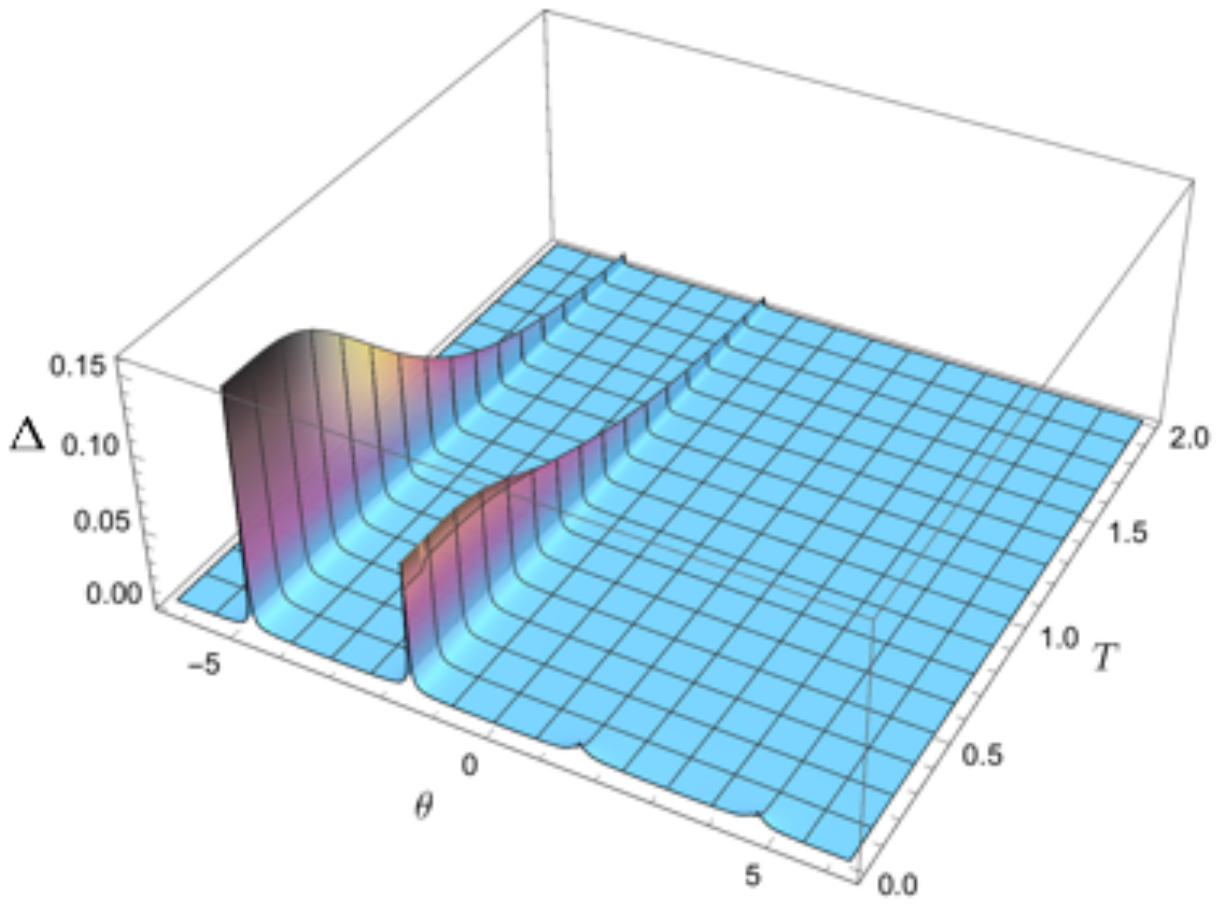}
\includegraphics[width=0.20\textwidth,height=0.15\textwidth]{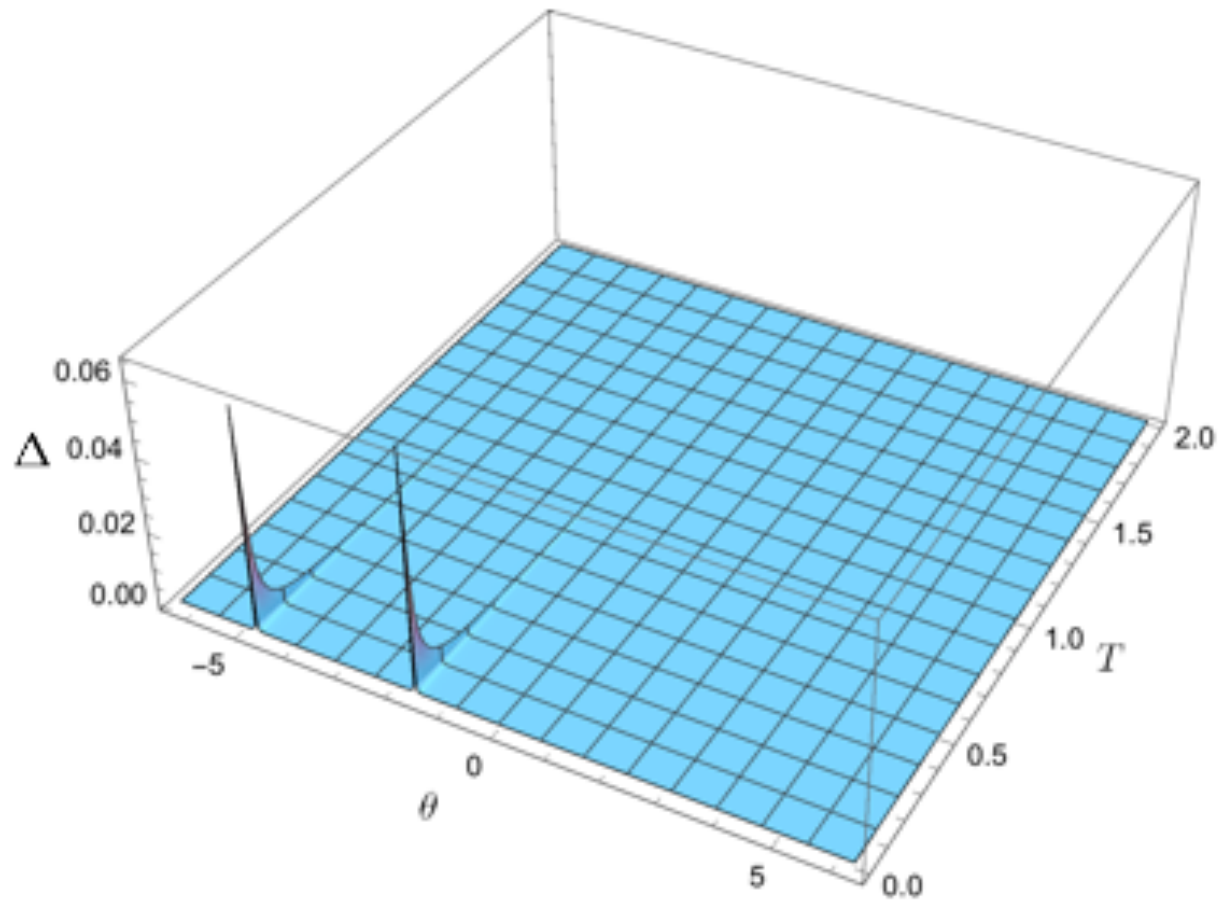}
\includegraphics[width=0.20\textwidth,height=0.15\textwidth]{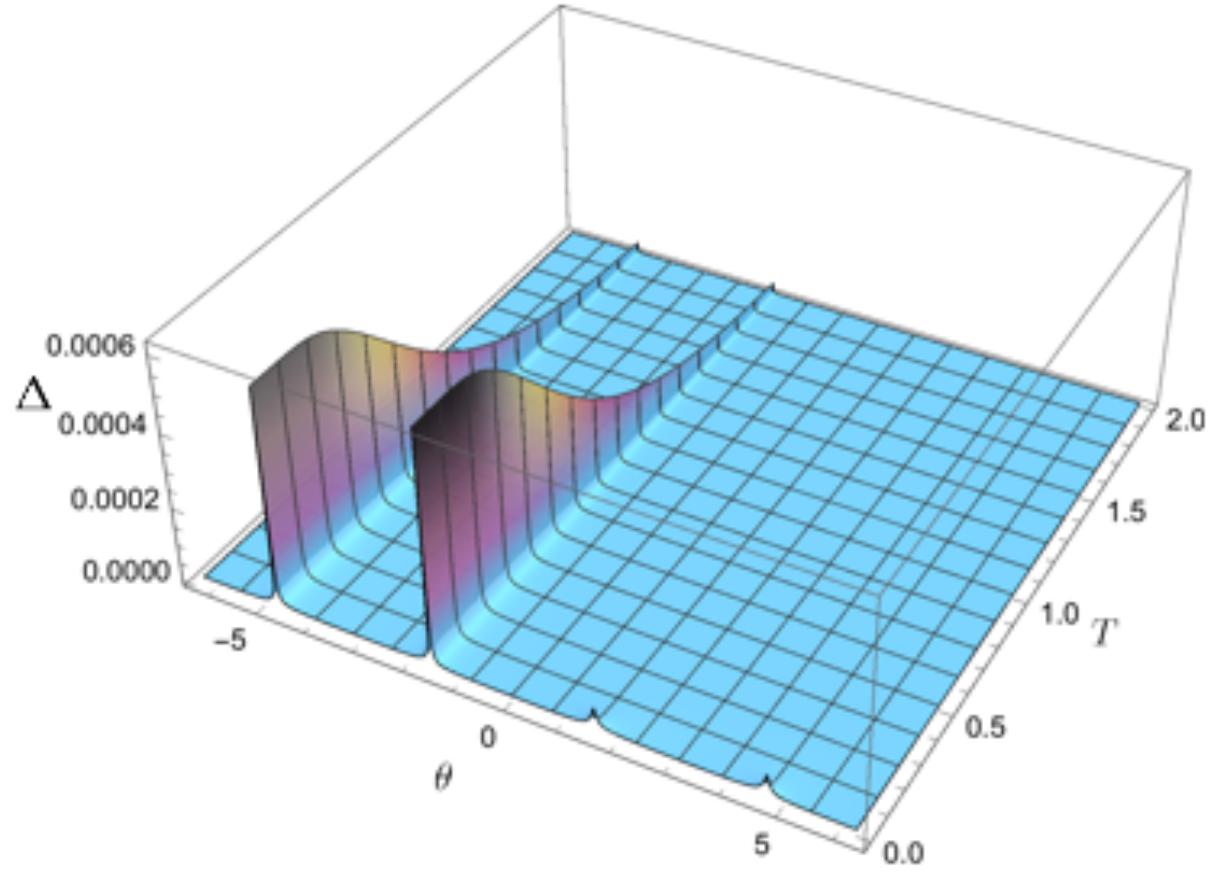}
\end{flushleft}
\end{minipage}
\begin{minipage}{1\textwidth}	
\caption{Fidelity (top) and $\Delta(\rho,\rho')$ (bottom) for the many-body states $\varrho^{(0)}$ and $\varrho^{(1)}$ and the single-particle states $\rho^{(0)}$ and $\rho^{(1)}$, for the BDI symmetry class. $\delta \theta =\theta'-\theta=0.01$ and $\delta T=T'-T=0.01$. The small step in the top middle-left plot is due to numerical instability (see Appendix 4).}
\label{fig:fidelity}
\end{minipage}
\end{figure}

 In our analysis, we study the overall quantum states over the parameter space denoted by $\vec{q}=(\beta^{-1},\theta)$, including temperature and the parameter of the Hamiltonian that drives the topological phase transition. We consider the fidelity $F$ and the quantity $\Delta$ between two states $\rho$ and $\rho'$ separated in the parameters' space  by an ``infinitesimal'' displacement: namely $F(\rho,\rho')$ and $\Delta(\rho,\rho')$, where prime denotes the ``infinitesimally'' close parameters.  We can consider three different cases. In the first case, which is the most general, we probe the system with respect to both the parameter $\theta$ and the temperature $T$ on the same time, that is:
 $$\rho'=\rho(\vec{q}\prime)=\rho(\vec{q}+\delta \vec{q}),$$ with $\delta \vec{q}=(\delta\theta,\delta T)$ and $||\delta \vec{q}||<<||\vec{q}||,$ since $|\delta \theta|<<|\theta|$ and $|\delta T|<<|T|.$
 The analytic derivation of the expressions for $F$ and $\Delta$ of the BG states considered can be found in Appendix 2. The other two cases occur when we wish to probe the system with respect to the parameter of the Hamiltonian and the temperature separately, these are for $\delta \vec{q}=(\delta\theta,0)$ and $\delta \vec{q}=(0,\delta T)$, respectively. The quantitative results for the representative of the class BDI are given in Fig.~\ref{fig:fidelity}. The results for the representative of the class AIII are qualitatively similar and, thus, we present them in Appendix 3. For both classes, the quantitative results that we present in this work are obtained in the most general case, where we probe the system with respect to $\theta$ and $T$ on the same time, i.e., $\delta \vec{q}=(\delta\theta,\delta T)=(0.01,0.01).$ In what follows we also comment on the results obtained for the other two cases $\delta \vec{q}=(\delta\theta,0)=(0.01,0)$ and $\delta \vec{q}=(0,\delta T)=(0,0.01)$, however we do not present them, as they are qualitatively similar.

A unique feature of periodically driven systems and their corresponding effective Hamiltonians is that both energies $E_k = 0$ and $E_k = \pi$ correspond to a closed gap (the difference between the two energy levels $\pm E_k$ becomes zero modulo $2\pi$). This special feature of QWs yields a surprising result in our analysis. In our study, we observe a different behaviour of the gap closing points with temperature, depending on whether they correspond to $E_k = 0$ (the two points with $\theta > 0$) or $E_k = \pi$ energy (the two points with $\theta < 0$). 
 Whenever the gap closes, the vector $\vec{n}$ is ill-defined (for its specific form, see~\cite{kit:rud:ber:dem:10}), but the behaviour of $F$ and $\Delta$ will be different due to their dependence on the entire energy spectrum. For the case of $E_k = 0$, they signal two isolated zero-temperature points of quantum phase transitions, corresponding to $\theta = \pi /2,\ 3\pi /2$ (in the case of $\Delta$ we notice small, but still present, peaks at these points). As the temperature increases, they are no longer signalling a phase transition and this is due to the dependence on the hyperbolic sine of $E_k$, which vanishes for $E_k=0$ (see the respective formulae in Appendix 2), thus eliminating the $\vec n\cdot \vec n'$ term carrying the relevant topological features. The significance of these points of the zero-temperature quantum phase transitions on the system in the low-temperature regime, and the existence of possible crossovers, remain as open questions and require further investigation. In contrast to that, for the $E_k=\pi$ gap closing points, the phase transition lines survive with temperature, hence revealing a ``finite-temperature quantum phase transition'' (a phase transition occurring at finite temperature, driven solely by the Hamiltonian's parameter(s), and not the temperature). Again, this can be understood through the dependence on $E_k$ via hyperbolic functions which take finite values for $E_k=\pi$, thus maintaining the dependence on $\vec{n}\cdot\vec{n}'$.
 
 Notice that for $\rho^{(0)}$ the qualitative behaviour is different from the other three types of states: the fidelity does not drop for $T=0$ and $\theta = \pi /2, 3\pi /2$, while it does for two new $T=0$ points at $\theta = \pm\pi$. The first difference is due to the fact that the zero temperature limit of $\rho^{(0)}$ projects only onto $M$ given by the points of minimum energy $-E_k=-\pi$, see Eq.~\eqref{eq:rho0_lim}. Thus, both quantities do not see the critical momentum at which the gap closes at zero energy, which is above the lowest mode. The abrupt change of the fidelity at the points where $\theta=\pm \pi$ is the consequence of the enhanced zero-temperature ground state distinguishability due to the fact that $E_k$ becomes constant and independent of $k$, i.e.,  the zero-temperature state projects onto the whole Brillouin zone ($M(\theta = \pm\pi) = \mathcal B$). Notice also that in the respective plot for $\Delta$, the absence of peaks at $\theta = \pm\pi$ is consistent with the fact that the Uhlmann factor quantifies the change of eigenvectors, while the presence of two peaks at the corresponding fidelity plot is due to the flattening of the spectrum, i.e., solely to the change of eigenvalues.
 
Let us, now, comment on the results for the fidelity and the quantity $\Delta$, in the case where we probe the system with respect to the parameter of the Hamiltonian and the temperature, separately. In the first case, where $\delta\theta=0.01$ and $\delta T=0$ ($\delta\vec{q}=(0.01,0)$), the results are qualitatively the same as the ones presented in Figure~\ref{fig:fidelity} for $\delta\vec{q}=(0.01,0.01).$ In the second case, where we probe the system only with respect to temperature, that is $\delta\theta=0, \delta T=0.01$ or more concisely $\delta\vec{q}=(0,0.01),$ the results for the fidelity are qualitatively the same as the ones for $\delta\vec{q}=(0.01,0.01),$ while the results for the quantity $\Delta$ are different. In particular, we obtain that $\Delta$ is equal to zero everywhere. This triviality is due to the fact that $\Delta$ quantifies the difference between the eigenbases of $\rho$ and $\rho'$, as mentioned before and in this case that we only change the temperature and not the parameter of the Hamiltonian, the eigenbasis remains the same. On the other hand, the fidelity drops at the points of the phase transition, since it is also sensitive to changes in the spectrum.

We proceed to the results regarding edge states. The bulk-to-boundary principle predicts their existence on the boundary between distinct topological phases. These states are symmetry protected, i.e., they are robust against perturbations of the Hamiltonian which respect the symmetries of the system. In the case of pure states, QWs realise the aforementioned principle as shown in~\cite{kit:rud:ber:dem:10}. To this end, the authors introduce a spatial dependence on the parameter $\theta$ of the Hamiltonian in such a way that a phase boundary is created (see also Appendix 1 for a brief description of the method). The edge states are then observed by evolving the walk \textit{in time} from the initial state localised at the phase boundary: the probability to find the system in the initial position keeps being (considerably) higher than for the rest of the line. Instead, in our case, we are interested in probing the robustness of the edge states against temperature. Therefore, we study an ensemble of the BG type which corresponds to a stationary state of the split-step QW realising the BDI class representative, with the position-dependent coin operation, parameterised by its temperature $\beta^{-1}$. This stationary state appears naturally if one looks at the von Neumann evolution equation for the density matrix and imposes the stationarity condition. By diagonalising $H_{\text{eff}}$ we obtain the localised states which can either have quasienergy $0 \text{ or }\pi$. Here, we choose a different order for the energy spectrum, by promoting the edge states with energies $E=\pi$ and $E = 0$ to be ground states, so that they survive at the zero-temperature limit. 

\begin{figure}[h!]
\center
\includegraphics[width=0.5\textwidth,height=0.25\textwidth]{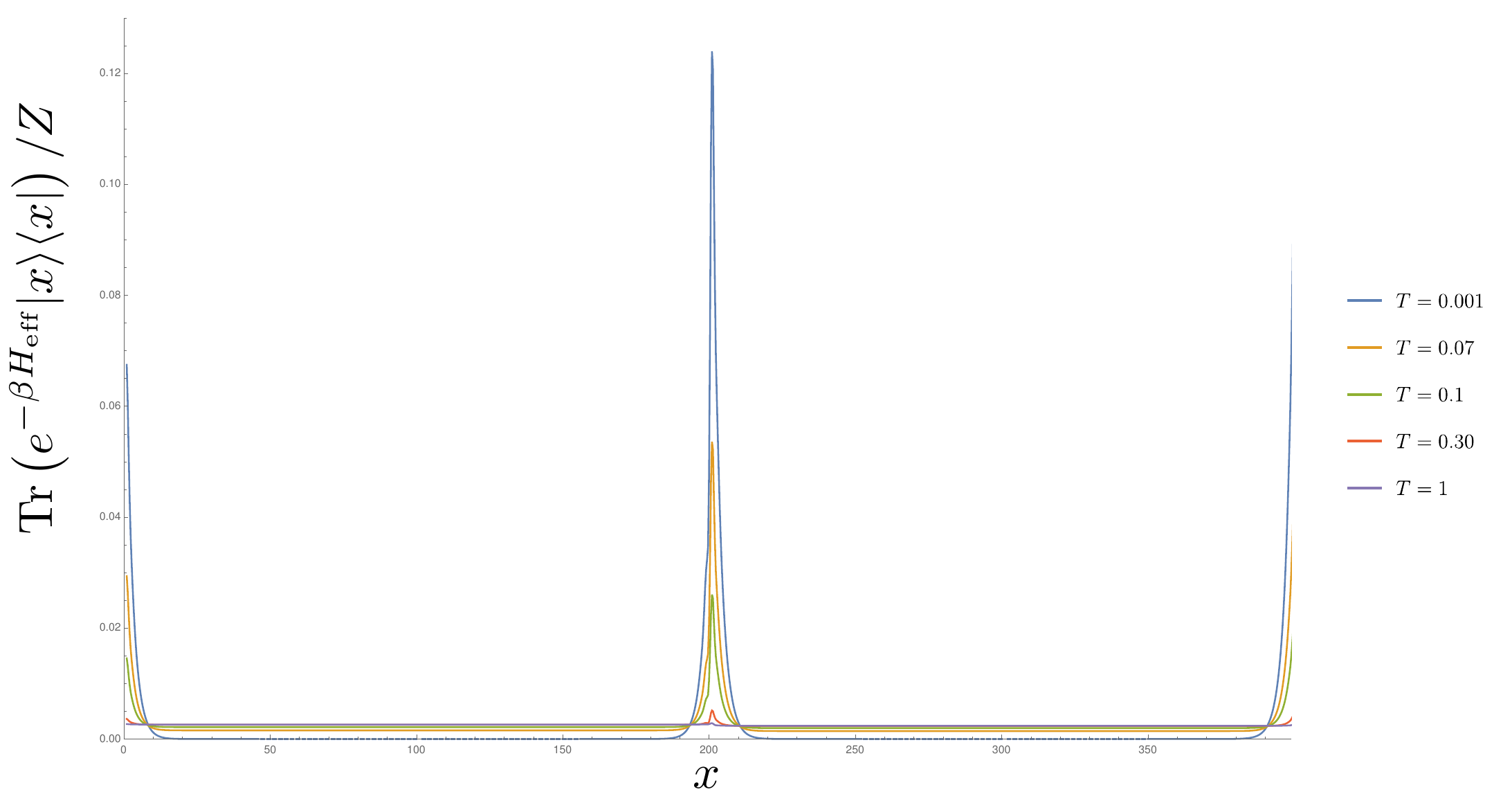}
\caption{
Position probability distribution of the QW as a function of the sites, $\tr (e^{-H/T}\ket{x}\bra{x})/Z$. The Hamiltonian $H$ is obtained by varying $\theta$ along $x$ through a step-like function~\cite{kit:rud:ber:dem:10}. The domain wall is centred in the middle of the line. Periodic boundary conditions are taken, hence the edge state at the boundary.}
\label{fig: edge}
\end{figure}

At $T=0$, the edge state has the major contribution to the probability distribution. As temperature increases, the edge states are smeared out, see Fig.~\ref{fig: edge}. This was also observed in~\cite{riv:viy:del:13}, by studying Chern values.
Since the presence of an edge state is a clear manifestation of the topological order of a system at $T=0$ then, the fact that it does not disappear for high temperatures, but it is rather smeared out, shows explicitly that the topological nature of the system is preserved in agreement with the fidelity analysis for the $E_k=\pi$ gap-closing points.

Our fidelity analysis reveals that for $T>0$ there exist no thermal phase transitions (i.e., no temperature-driven phase transitions), as also shown in~\cite{mer:vla:pau:vie:17} for paradigmatic models of topological insulators and superconductors. However, here we observe finite temperature \emph{parameter} $\theta$-driven phase transitions, a property inherited from the temporal periodicity of the single-particle QW evolution. The non-existence of thermal phase transitions is further confirmed by the behaviour of the edge states, which are gradually smeared out as temperature increases, as also pointed out in~\cite{viy:riv:del:14}. Note that the edge states studied in the current paper are probed through a different method from the one used in~\cite{mer:vla:pau:vie:17}. There, a small chemical potential was introduced in such a way that the zero-temperature behaviour of the many-body density operator is the projector onto the Fermi sea. Consequently, the existence of the edge states was signalled by the abrupt change in the number operator on the sites where they appeared.

\section{Conclusions}
\label{Conclusions and Outlook}
We derived analytic expressions for the fidelity and the quantity $\Delta$ between two BG states for QW representatives of topological phases for the chiral symmetric classes BDI and AIII and their many-body counterparts. For the systems considered, the fidelity is detecting the points where the Bloch vector is ill-defined, which corresponds to the closure of the quasienergy gap. Since in our case the phase diagram is such that the closure of the gap always means the change from a trivial phase to a nontrivial one, and vice versa, the fidelity is capturing the topological phase transitions. Our results show the absence of temperature-driven phase transitions in two ways: through the fidelity analysis and through the behaviour of the edge states appearing on the phase boundary. In addition, the analysis of the fidelity and the Uhlmann connection, through the quantity $\Delta$, shows the existence of finite-temperature transitions driven solely by the Hamiltonian's parameter $\theta$, as a consequence of the temporal periodicity of the QW evolution. We would also like to point out that, providing that one of the goals of the paper is to study the finite-temperature behaviour of topological order in realistic many-body systems, the fact that the behaviour of the single-particle BG states is consistent with that of their many-body counterparts, shows that the analysis of the former can be a useful mathematical tool in the study of the latter.

Finally, we point out possible future lines of research. First, the same study could be applied to the rest of the symmetry classes in 1D and 2D using the protocols introduced in~\cite{kit:rud:ber:dem:10}. Further analysis of realistic noise effects that can give rise to BG states, in particular the single-particle ones, presents an interesting line of research (for a partial answer to this question, see also a recent study of the effects of thermal noise onto single-particle factor states of a topological insulator~\cite{riv:viy:del:13}). The above study could also be conducted for 2D QWs, as well as in the realm of multi-particle QWs.\\

The authors thank Sonja Barkhofen for providing clarifications regarding the experimental realisation presented in~\cite{bar_exp:16}.  
B. M. and C. V. acknowledge the support from DP-PMI and FCT (Portugal) through the grants SFRH/BD/52244/2013 and PD/BD/52652/2014, respectively. B. M. acknowledges the support of Physics of Information and Quantum Technologies Group.  C. V. and N. P. acknowledge the support of SQIG -- Security and Quantum Information Group and UID/EEA/50008/2013. N. P. acknowledges the IT project QbigD funded by FCT PEst-OE/EEI/LA0008/2013. V. R. V. acknowledges support from FCT (Portugal) through Grant UID/CTM/04540/2013.

\section*{Appendix}
\subsection*{1. Quantum walk protocols}
In this section, we briefly describe the quantum walk protocols we used for the simulation of the two chiral-symmetric classes BDI and AIII, as introduced in~\cite{kit:rud:ber:dem:10}. The Hilbert space of a one-dimensional discrete-time quantum walk (DTQW) is the tensor product of the position Hilbert space $H_p = \text{span} \{ \ket{i}_p | i\in \mathbb Z \}$, where $i$ refer to the sites of an infinite line, with the two-dimensional coin Hilbert space $H_c = \text{span} \{ \ket{0}_c , \ket{1}_c \}$ (for simplicity, in the following we drop the subscripts $p$ and $c$ for the states). At each step, the walker is moving to the left or to the right on the line depending on its coin state. In a standard DTQW on a line, each step of the walk is given by the following unitary operator:
\begin{align}
	U=T\cdot(I_p\otimes R),
\end{align}
where $R \in U(2)$ is a rotation in the 2-dimensional coin space, and $T$ is the unitary operator describing the conditional shift of the walker (moving one position to the right when the coin state is $\ket 0$, and to the left if it is $\ket 1$):
\begin{align}
T=\sum_x\big(\ket{x+1}\bra{x}\otimes\ket{0}\bra{0}+\ket{x-1}\bra{x}\otimes\ket{1}\bra{1}\big).
\end{align}
In~\cite{kit:rud:ber:dem:10} the authors show that this quantum walk simulates the SSH (Su-Schrieffer-Heeger) non-trivial topological phase with winding number 1. The SSH model is a representative of the BDI symmetry class, studied in the Letter. In order to study the behaviour of edge states, which appear on the boundary between two distinct topological phases, the authors in~\cite{kit:rud:ber:dem:10} introduce the so-called \textit{split-step} quantum walk. As the name suggests, each step of the walk is split in two parts, each having the structure analogous to that of a standard DTQW,
\begin{align}
U_{ss}=T_{1}R(\theta_2)T_{0}R(\theta_1),
\label{eq:split}
\end{align}
where $R(\theta_i)$ represent rotation for the angle $\theta_i$, along a fixed common axis $\vec \alpha$ (which, for simplicity, we omit to explicitly indicate), and
\begin{eqnarray}T_{0} & = & \sum_x\big(\ket{x+1}\bra{x}\otimes\ket{0}\bra{0}+\ket{x}\bra{x}\otimes\ket{1}\bra{1}\big), \nonumber \\
T_{1} & = & \sum_x\big(\ket{x-1}\bra{x}\otimes\ket{1}\bra{1}+\ket{x}\bra{x}\otimes\ket{0}\bra{0}\big).
\end{eqnarray}
For different values of the parameters $\theta_1$ and $\theta_2$ this protocol is shown to realise both trivial ($\nu=0$) and non-trivial ($\nu=1$) SSH topological phases. We choose $\theta_1$ to be the same for both topological phases, while $\theta_2$ is different. The closure of the gap implies a change of phase, in our case from a trivial to a non-trivial one and vice versa, so that by varying the value of $\theta_2$, thus changing the energy spectrum, we are able to close the gap, and have a phase transition. That means that by making $\theta_2$ change along the line, $\theta_2=\theta_2(x)$, we can create a phase boundary on the site $x_0$ of the line, where $\theta_2$ changes. Specifically, $\theta_2(x)$ is chosen to be $\theta_2(x) = \theta_2^{(t)}H(x_0-x) + \theta_2^{(n)}H(x-x_0)$, where $H(x)$ is the Heaviside step function, and $\theta_2^{(t/n)}$ are the values of the angle for the trivial/non-trivial phase. Starting the quantum walk on this site (where the phase boundary is), permits us to detect the edge states, since even after many steps of the walk, there is still a large probability to find the walker at the origin, as a consequence of the overlap between the initial state with the edge state that is localised there~\cite{kit:rud:ber:dem:10,kit:12}.

In order to simulate different symmetry classes (for the classification of topological insulators according to their symmetries, see~\cite{sch:ryu:fur:lud:08,kit:09}), we further modify this basic split-step protocol. By changing the spin rotation axis, we manage to break Time-Reversal (TRS), and Particle-Hole (PHS) Symmetry, while maintaining Chiral Symmetry (CS), given by the operators $\mathcal T, \mathcal P$ and $\Gamma$, respectively. That leads us to a split-step protocol that simulates the symmetry class AIII. 

The chiral symmetry operators are, for fixed $\theta_1$, given as follows~\cite{kit:rud:ber:dem:10,kit:12}: 
\begin{eqnarray}\Gamma_{\theta_1}^y&=&\exp(-i\pi \vec{A}_{\theta_1}^y\cdot\vec\sigma/2) \nonumber\\
\Gamma_{\theta_1}^{\alpha}&=&\exp(-i\pi \vec{A}_{\theta_1}^{\alpha}\cdot\vec\sigma/2),\end{eqnarray}
for BDI and AIII, respectively. For each chiral class representative, this operator is determined by the vector $\vec{A}_{\theta_1}^{y/\alpha}$, perpendicular to unit vectors $y=[0,1,0]$ and $\alpha=1/\sqrt{2}[0,1,1]$, and given by $\vec{A}_{\theta_1}^y=[\cos(\theta_1/2),0,\sin(\theta_1/2)]$ (for BDI class) and $\vec{A}_{\theta_1}^{\alpha}=[\cos(\theta_1/2),-1/\sqrt{2}\sin(\theta_1/2),1/\sqrt{2}\sin(\theta_1/2)]$ (for AIII class), respectively. Thus, operators $\Gamma_{\theta_1}^{y/\alpha}$ rotate $\vec{n}_{\theta_1}(k)$ by the angle $\pi$ in a plane orthogonal to $\vec{A}_{\theta_1}^{y/\alpha}$. In class BDI, PHS is given by the operator
\begin{align}
\mathcal{P}=	\mathcal{K},
\end{align}
where $\mathcal{K}$ denotes the complex conjugation operator, and TRS by the operator:
\begin{align}
\mathcal{T}=	\Gamma_{\theta_1}^y\mathcal{P}.
\end{align}
In Table I, we present the aforementioned 1D chiral classes, their symmetries, the quantum walk protocol that simulates them and the values of the parameters for each distinct topological phase, characterised by a winding number.
\begin{center}
\begin{table}[h!]\center
\begin{small}
\begin{tabular}{c c c c c c c } \hline\hline
Class &TRS & PHS &CS&Protocol&Parameters&$\nu$\\\hline
BDI&$\mathcal{T}^2=1$&$\mathcal{P}^2=1$&$(\Gamma_{\theta_1}^y)^2=1$&$T_1R_y(\theta_2)T_0R_y(\theta_1)$&\raisebox{2ex}{$\theta_1=-\pi/2,\theta_2=3\pi/4$}&\raisebox{2ex}{$\nu=0$}\\
&&&&&$\theta_1=-\pi/2,\theta_2=\pi/4$&$\nu=1$\\[1ex]\hline
\raisebox{0.5ex}{AIII}&\raisebox{0.5ex}{Absent}&\raisebox{0.5ex}{Absent}&\raisebox{0.5ex}{$(\Gamma_{\theta_1}^{\alpha})^2=1$}&&\raisebox{4ex}{$\theta_1=\pi/2,\theta_2=3\pi/4$}&\raisebox{4ex}{$\nu=0$}\\[-3ex]
&&&&\raisebox{2ex}{$T_1R_{\alpha}(\theta_2)T_0R_{\alpha}(\theta_1)$}&$\theta_1=\pi/2,\theta_2=\pi/4$&$\nu=1$\\\hline\hline
\end{tabular}
\caption{Chiral Symmetric Classes in 1D and the respective quantum walk protocols. The values of the parameters  $\theta_1$ and $\theta_2$ that correspond to distinct topological phases are shown, as well as the respective winding numbers $\nu$ of each phase.}
\end{small}
\end{table}
\end{center}

\subsection*{2. Analytic derivation of the closed expression for the fidelity}
Given two thermal unnormalised states $\rho=\exp(-\beta H)$ and $\rho'=\exp(-\beta'H')$ the fidelity is given by
\begin{align}
F(\rho,\rho')=\text{Tr}\sqrt{\sqrt{\rho}\rho'\sqrt{\rho}}.
\end{align}
At the end of the calculation one must, of course, normalise the expression appropriately. In order to find closed expressions for the four possible density matrices we present in the main text, we will find $e^C$, such that
\begin{align}
e^{A}e^{B}e^{A}=e^{C},
\end{align}
for $A=-\beta H$, $B=-\beta'H'$ and, then we will take the square root of the result. Equivalently, the previous equation can be written as
\begin{align}
e^{A}e^{B}=e^{C}e^{-A}.
\label{eq:1}
\end{align}
By considering the Hamiltonians to be of the form
 $\q{h}$, we can write
\begin{align}
e^{A}=a_0+\q{a}, \nonumber \\
e^{B}=b_0+\q{b}, \nonumber \\
e^{C}=c_0+\q{c},
\end{align}
where the coefficients are real and they satisfy the constraint:
\begin{align}
\label{eq:c1}
 1=\det e^{A}=a_{0}^2-\vec{a}^2
\end{align}
The same constraint holds for the determinants of $B$ and $C$.
We expand the LHS and the RHS of Eq.\eqref{eq:1} and, by collecting the terms in $1$, $\vec{\sigma}$ and $i\vec{\sigma}$, we get the following system of linear equations for $c_0$ and $\vec{c}$,
\begin{align}
\label{Eq:LS1}
\begin{cases}
& a_0 b_0+\vec{a}\cdot\vec{b}- a_0 c_0 +\vec{a}\cdot\vec{c}=0\\
& a_0\vec{b}+b_0\vec{a}+\vec{a}c_0-a_0\vec{c}=0\\
& \vec{a}\times (\vec{b}-\vec{c})=0
\end{cases}.
\end{align}
The third equation has the solution $\vec{c}=\vec{b}+\lambda\vec{a}$, where $\lambda$ is a real number. This means that the solution depends only on two real parameters: $c_0$ and $\lambda$. After some straightforward algebra, we obtain the solution:
\begin{align}
\left[\begin{array}{c}
c_0\\
\lambda	
\end{array}
\right]=\left[\begin{array}{c}
(2 a_0 ^2-1) b_0+2 a_0\vec{a}\cdot\vec{b}\\
2(a_0 b_0+\vec{a}\cdot\vec{b})
\end{array}
\right].
\end{align} 
Next, we set $A=-\beta H/2\equiv-\xi \vec{x}\cdot \vec\sigma/2$ and $B=-\beta 'H'\equiv-\zeta \vec{y}\cdot \vec\sigma$, with $\vec{x}^2=\vec{y}^2=1$ and $\xi,\zeta$ are real parameters, meaning,
\begin{align}
a_0=\cosh(\xi/2) \text{ and } \vec{a}=-\sinh(\xi/2) \vec{x},\\
b_0=\cosh(\zeta) \text{ and } \vec{b}=-\sinh(\zeta) \vec{y}.	
\end{align}
If we write $C=\rho \vec{z}\cdot \vec \sigma$ (which follows from the property of the product of special matrices) we get,
\begin{align}
c_0 =\cosh(\rho)=\cosh(\xi)\cosh(\zeta)+\sinh(\xi)\sinh(\zeta)\vec{x}\cdot\vec{y}.
\end{align}
Using the formula $\cosh(\rho/2)=\sqrt{(1+\cosh(\rho))/2}$, we obtain,
\begin{align}
\tr(e^{C/2})=2\sqrt{\frac{(1+\cosh(\xi)\cosh(\zeta)+\sinh(\xi)\sinh(\zeta)\vec{x}\cdot\vec{y})}{2}},
\end{align}
and by substituting, we end up with the following formula:
\begin{align}
\tr(\sqrt{e^{-\beta H/2}e^{-\beta' H'}e^{-\beta H/2}})=2\sqrt{\frac{(1+\cosh(\beta \Delta )\cosh(\beta'\Delta')+\sinh(\beta\Delta)\sinh(\beta'\Delta')\vec{n}\cdot\vec{n}')}{2}}.
\end{align}
To be able to compute all the fidelities, we will just need the following expression relating the traces of quadratic many-body fermion Hamiltonians (preserving the number operator) and the single-particle sector Hamiltonian obtained by projection:
\begin{align}
\tr{e^{-\beta \mathcal{H}}}=\tr e^{-\beta \Psi^{\dagger}H\Psi}=\det(I+e^{-\beta H}).
\end{align}
From the previous results, it is straightforward to derive the following formulae for the fidelities concerning all the {\em normalised} states discussed in the main text:\\

\textbf{Single-Body}
\begin{align}
	F(\rho^{(0)},\rho'^{(0)})&=\frac{\sum_{k\in\mathcal{B}}\tr (e^{-C_k/2})}{\sqrt{\sum_{k\in\mathcal{B}}\tr (e^{-\beta H_k})\sum_{k\in\mathcal{B}}\tr (e^{-\beta'H'_k})}} \nonumber\\
	&=\frac{\sum_{k\in\mathcal{B}} \sqrt{2\left(1+\cosh(E_k/2T)\cosh(E'_k/2T')+\sinh(E_k/2T)\sinh(E'_k/2T')\vec{n}_k\cdot\vec{n}'_k\right)}}{\sqrt{\sum_{k\in\mathcal{B}} 2\cosh (E_k/2T)\sum_{k\in\mathcal{B}}2\cosh (E'_{k}/2T')}}\\
	\nonumber \\
	F(\rho^{(1)},\rho'^{(1)})&=\sum_{k\in\mathcal{B}}\frac{\tr (e^{-C_k/2})}{\sqrt{\tr (e^{-\beta H_k})\tr (e^{-\beta'H'_k})}} \nonumber\\
	&=\sum_{k\in\mathcal{B}} \frac{\sqrt{2\left(1+\cosh(E_k/2T)\cosh(E'_k/2T')+\sinh(E_k/2T)\sinh(E'_k/2T')\vec{n}_k\cdot\vec{n}'_k\right)}}{\sqrt{ 2\cosh (E_k/2T)2\cosh (E'_k/2T')}}
\end{align}

\textbf{Many-Body}
\begin{align}
	F(\varrho^{(0)},\varrho'^{(0)})&=\prod_{k\in\mathcal{B}}\frac{\tr(e^{-\mathcal{C}_k/2})}{\tr(e^{-(\beta \mathcal{H}_k}))\tr(e^{-\beta' \mathcal{H}'_k})}\nonumber\\
	&=\prod_{k\in\mathcal{B}}\frac{\det(I+e^{-C_k/2})}{\det^{1/2}(I+e^{-\beta H_k})\det^{1/2}(I+e^{-\beta' H'_k})}\nonumber\\
	&=\prod_{k\in\mathcal{B}}\frac{2+\sqrt{2\left(1+\cosh(E_k/2T)\cosh(E'_k/2T')+\sinh(E_k/2T)\sinh(E'_k/2T')\vec{n}_k\cdot\vec{n}'_k\right)}}{\sqrt{[2+ 2\cosh (E_k/2T)][2+2\cosh (E'_k/2T')]}}\\
	\nonumber\\
	F(\varrho^{(1)},\varrho'^{(1)})&=\prod_{k\in\mathcal{B}}\frac{\tr(e^{-(\mathcal{C}_k-\log(Z_k)n_k-\log(Z'_k)n_k)/2})}{\tr(e^{-(\beta \mathcal{H}_k-\log(Z_k)n_k)})\tr(e^{-(\beta' \mathcal{H}'_k-\log(Z'_k)n_k)})}\nonumber\\
	&=\prod_{k\in\mathcal{B}}\frac{\det(I+e^{-C_k/2-\log(Z_k)/2-\log(Z'_k)/2})}{\det^{1/2}(I+e^{-(\beta H_k-\log(Z_k))})\det^{1/2}(I+e^{-(\beta' H'_k-\log(Z'_k))})}\nonumber\\
	&=\prod_{k\in\mathcal{B}} \left(1+\frac{[2\left(1+\cosh(E_k/2T)\cosh(E'_k/2T')+\sinh(E_k/2T)\sinh(E'_k/2T')\vec{n}_k\cdot\vec{n}'_k\right)]}{\sqrt{(2\cosh (E_k/2T))(2\cosh (E'_k/2T'))}} \right. \nonumber\\
	&\left.+\frac{1}{(2\cosh (E_k/2T))(2\cosh (E'_k/2T'))}\right)\times\left(\sqrt{[2+ (2\cosh (E_k/2T))^{-2}][2+(2\cosh (E'_k/2T'))^{-2}]}\right)^{-1},
\end{align}
where the single-particle partition function for momentum $k$, $Z_k$, is given by 
\begin{align}
Z_k &=\tr e^{-H_k/T}=2\cosh (E_k/2T),
\end{align}
the matrix $C_k$ is such that $e^{-C_k}=e^{-\beta H_k/2}e^{-\beta' H'_k}e^{-\beta H_k/2}$ and $\mathcal{C}_k=\Psi_k^{\dagger}C_k\Psi_k$ is the corresponding many-body quadratic operator.

To compute the quantity $\Delta(\rho,\rho')$ we need, in addition, $\tr \sqrt{\rho}\sqrt{\rho'}$. The derivation is similar but simpler, hence we shall omit the proof. 
\subsection*{3. Results for the representative of AIII symmetry class}
\begin{figure}[h!]
\center
\begin{minipage}{1.22\textwidth}
\begin{flushleft}
\includegraphics[width=0.20\textwidth,height=0.15\textwidth]{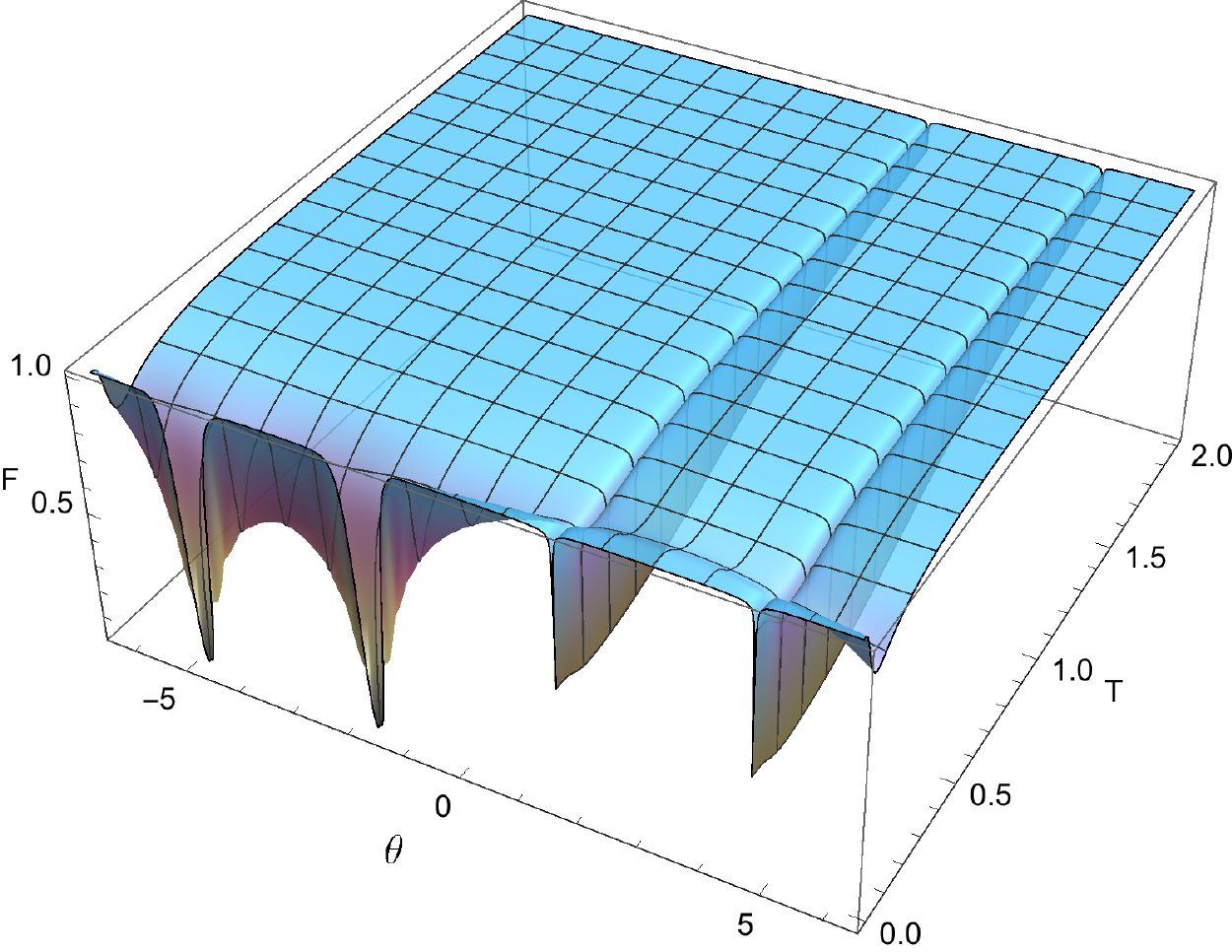}
\includegraphics[width=0.20\textwidth,height=0.15\textwidth]{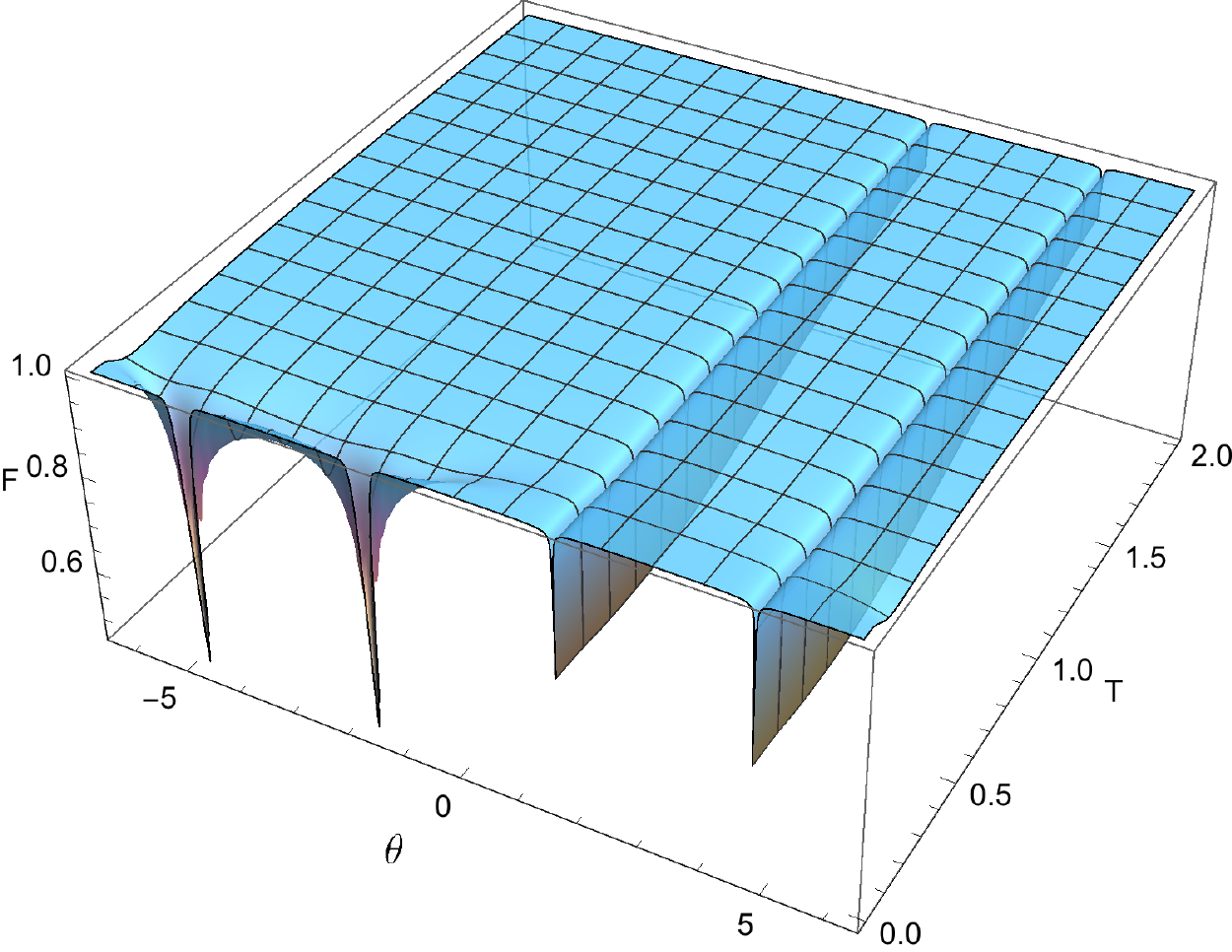}
\includegraphics[width=0.20\textwidth,height=0.15\textwidth]{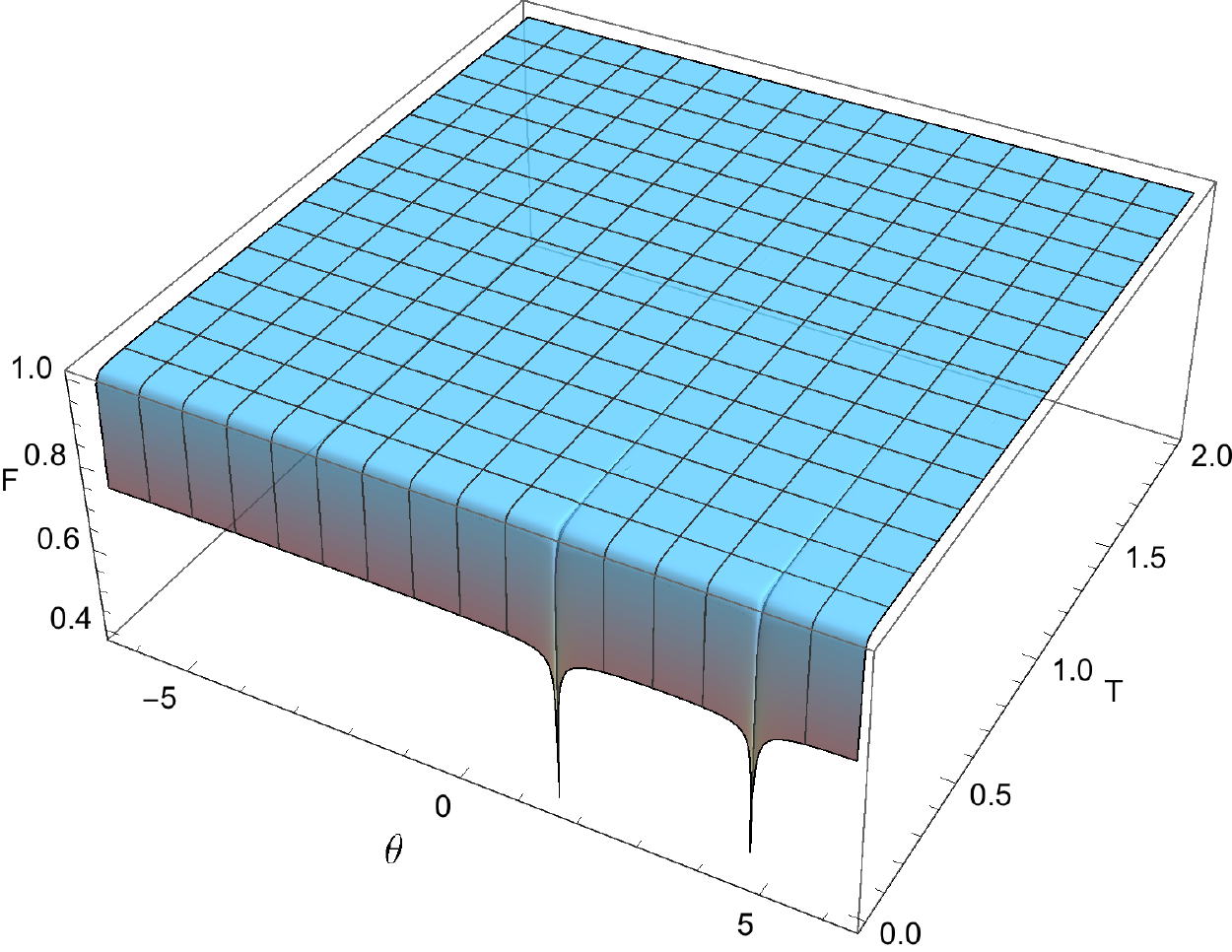}
\includegraphics[width=0.20\textwidth,height=0.15\textwidth]{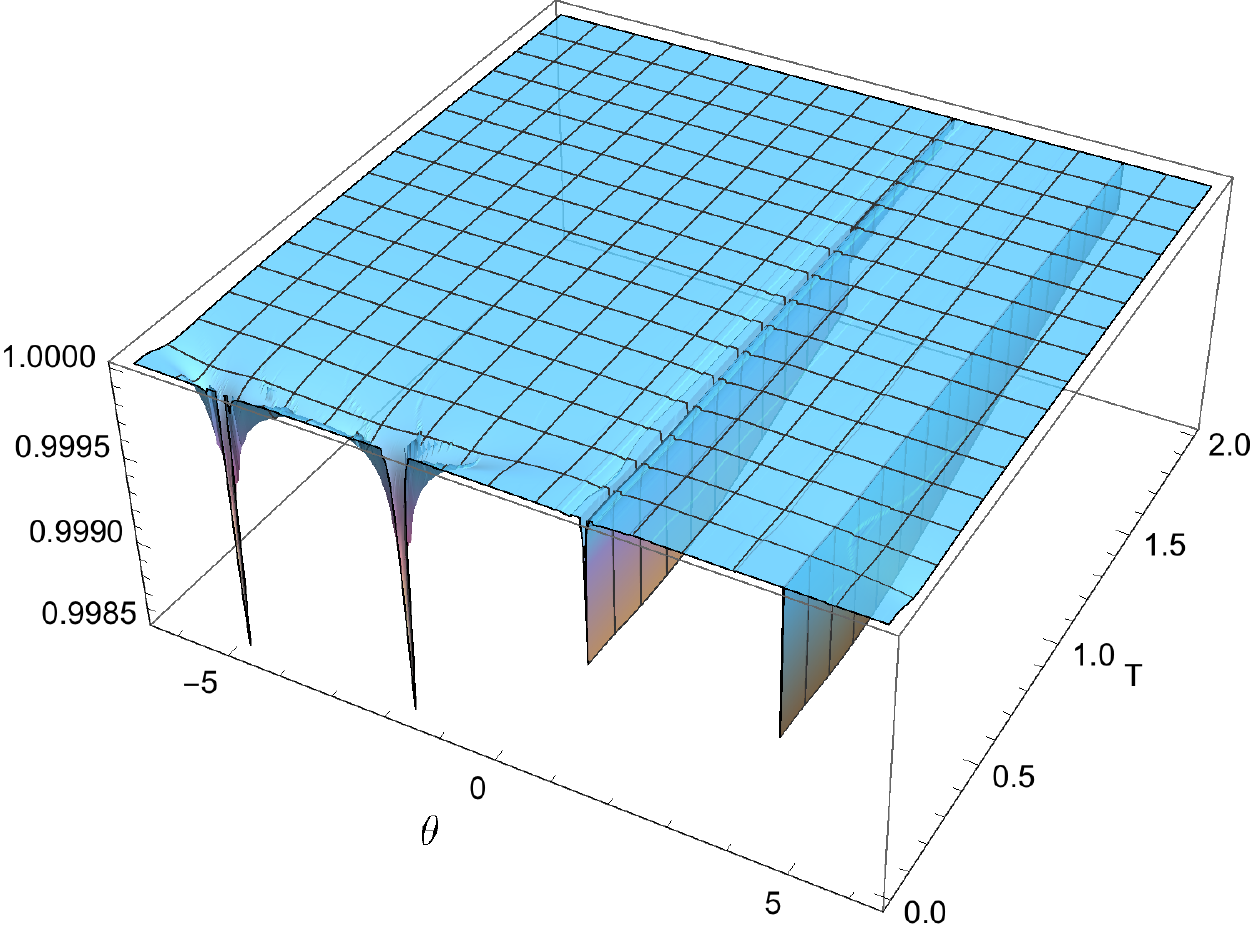}
\end{flushleft}
\end{minipage}
\begin{minipage}{1.22\textwidth}
\begin{flushleft}
\includegraphics[width=0.20\textwidth,height=0.15\textwidth]{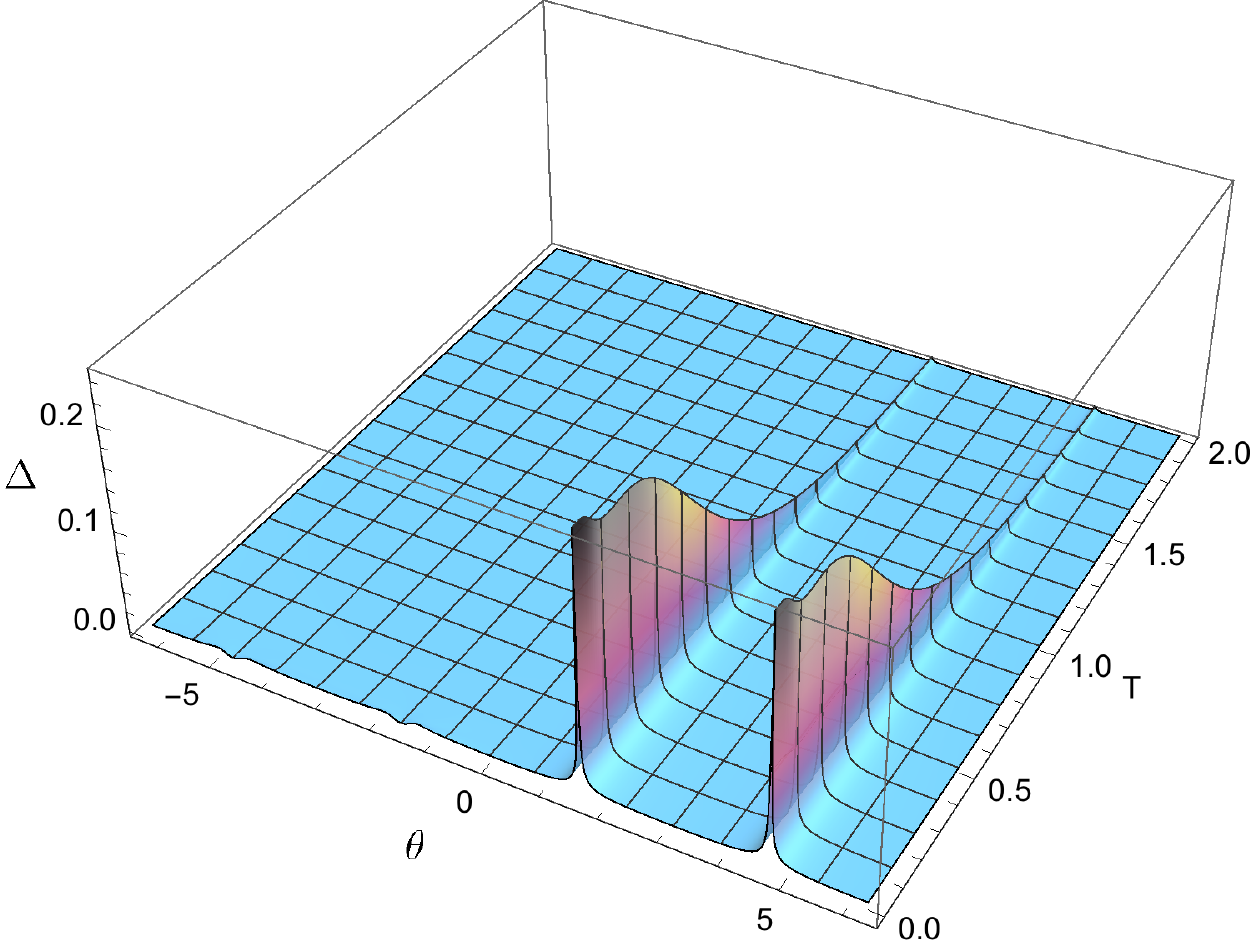}
\includegraphics[width=0.20\textwidth,height=0.15\textwidth]{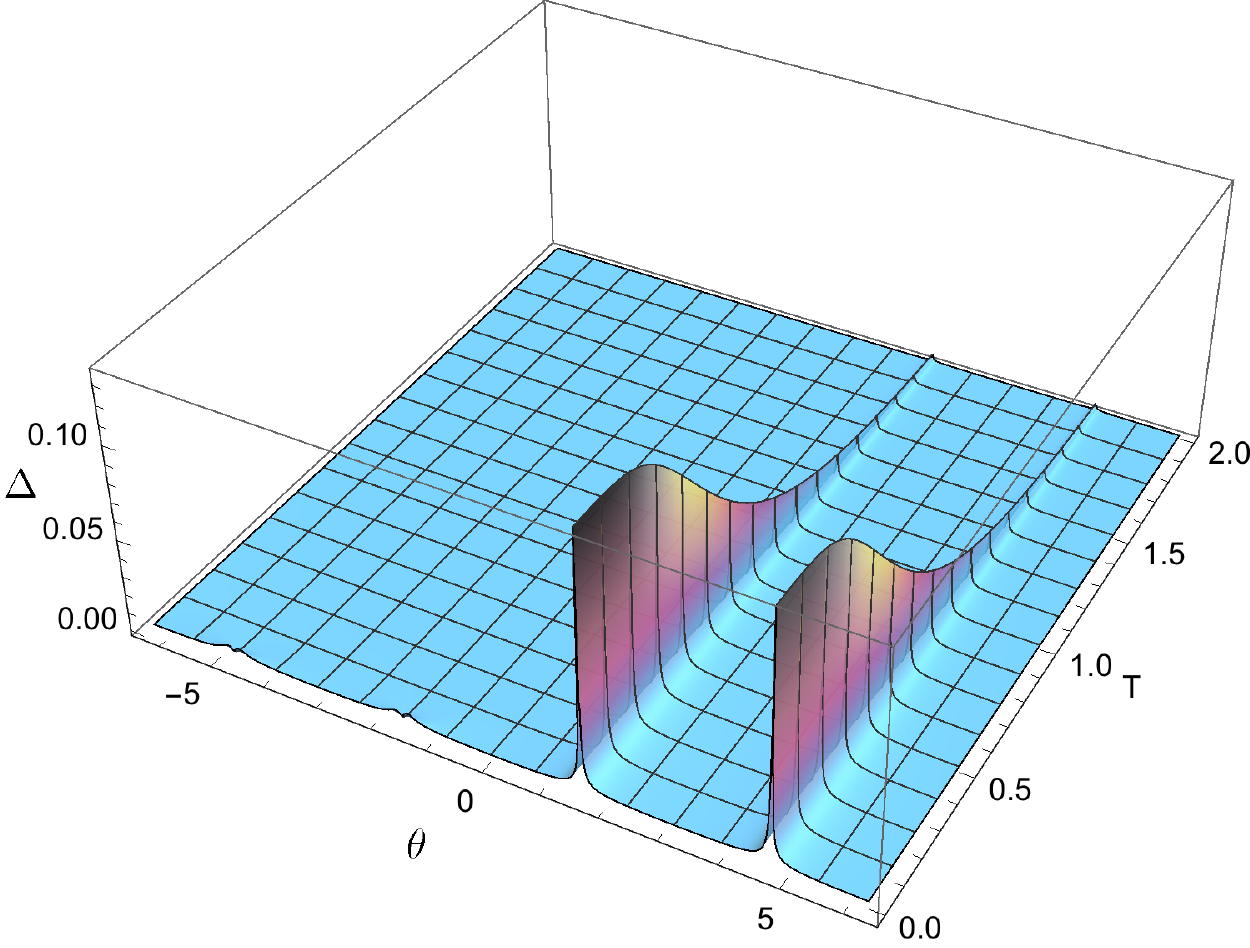}
\includegraphics[width=0.20\textwidth,height=0.15\textwidth]{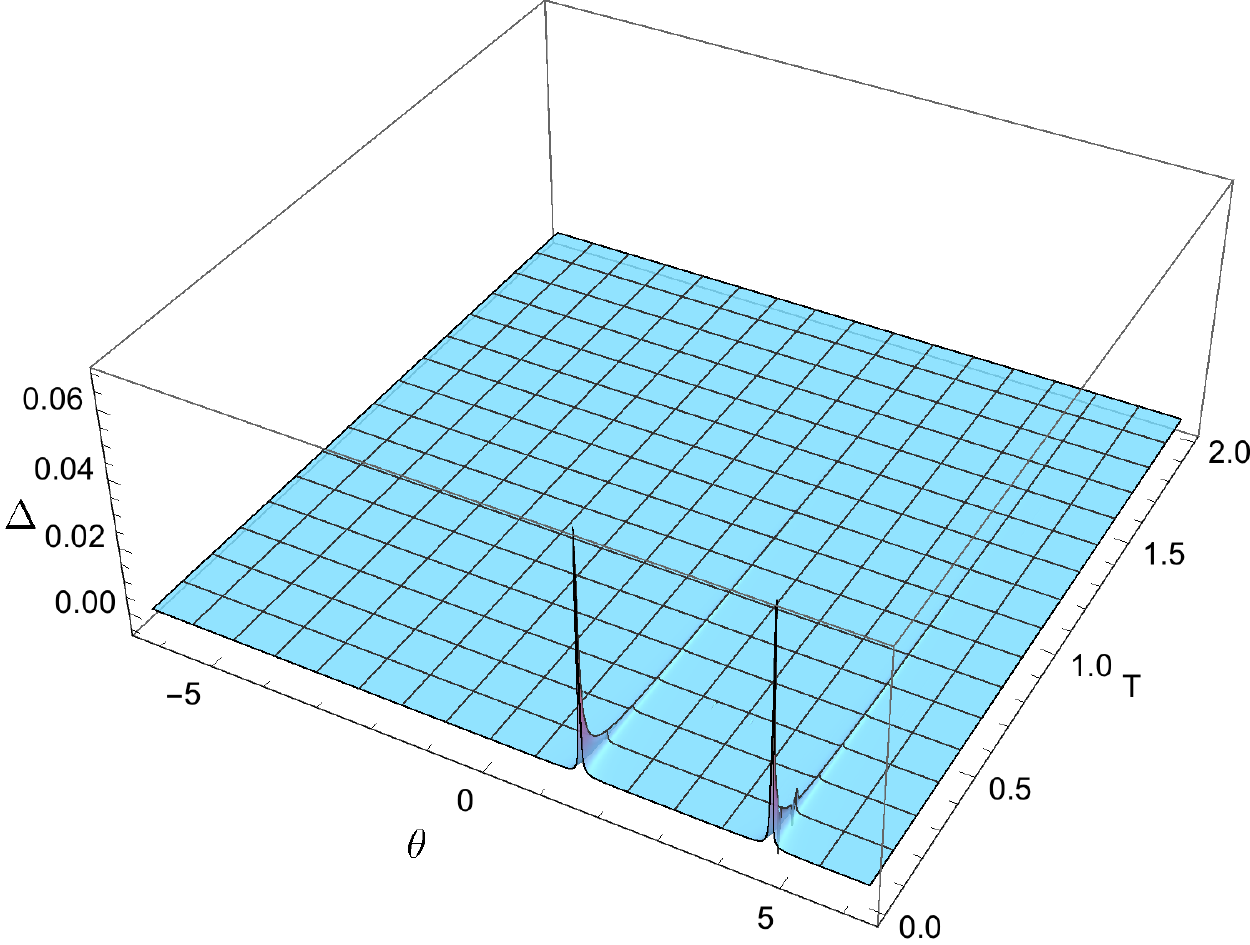}
\includegraphics[width=0.20\textwidth,height=0.15\textwidth]{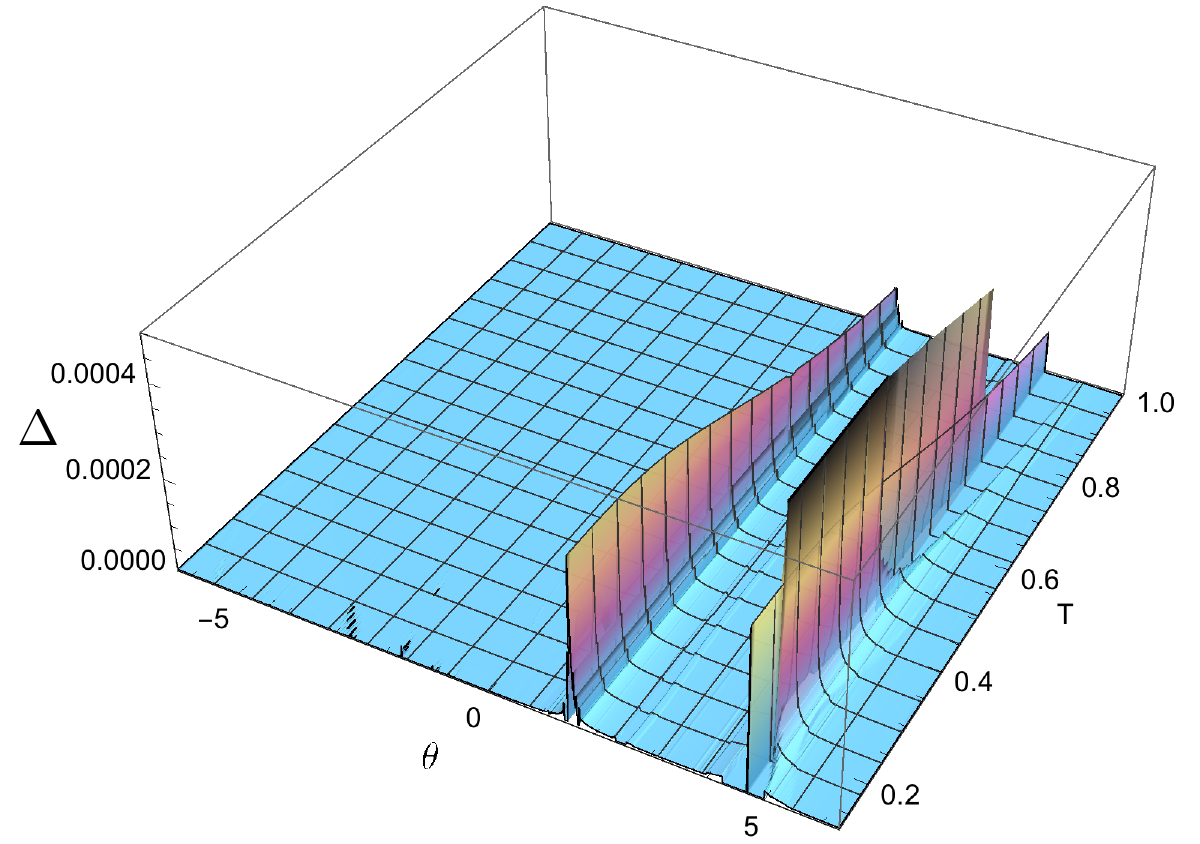}
\end{flushleft}
\end{minipage}
\begin{minipage}{1\textwidth}	
\caption{Fidelity (top) and $\Delta(\rho,\rho')$ (bottom) for the many-body states $\varrho^{(0)}$  and $\varrho^{(1)}$, and the single-particle states $\rho^{(0)}$ and $\rho^{(1)}$, for the AIII symmetry class. $\delta \theta =\theta'-\theta=0.01$ and $\delta T=T'-T=0.01$. In the case of the state $\rho^{(1)}$, the quantity $\Delta$ is highly oscillating for temperatures close to $0$ in the neighbourhood of the critical points, therefore we show the results for a range of temperatures where these numerical instabilities are less prominent.}
\label{fig:fidelityAIII}
\end{minipage}
\end{figure}

In Fig.~\ref{fig:fidelityAIII} we present the quantitative results for the fidelity and the quantity $\Delta$, in the case of the AIII symmetry class representative.
Here, we also observe the different behaviour of the gap closing points with temperature, depending on whether they correspond to $E_k = 0$ (in this case, these are the two points with $\theta < 0$) or $E_k = \pi$ energy (the two points with $\theta > 0$). 
While the results for the representatives of BDI and AIII classes are qualitatively similar, the zero-temperature fidelity corresponding to the $\rho^{(0)}$ states for the AIII representative only exhibits drops in the gap closing points ($-E_k=-\pi$), as the spectrum is always non-trivially $k$-dependent.
\\
\\

\begin{figure}[h!]
\center
\includegraphics[width=0.5\textwidth,height=0.25\textwidth]{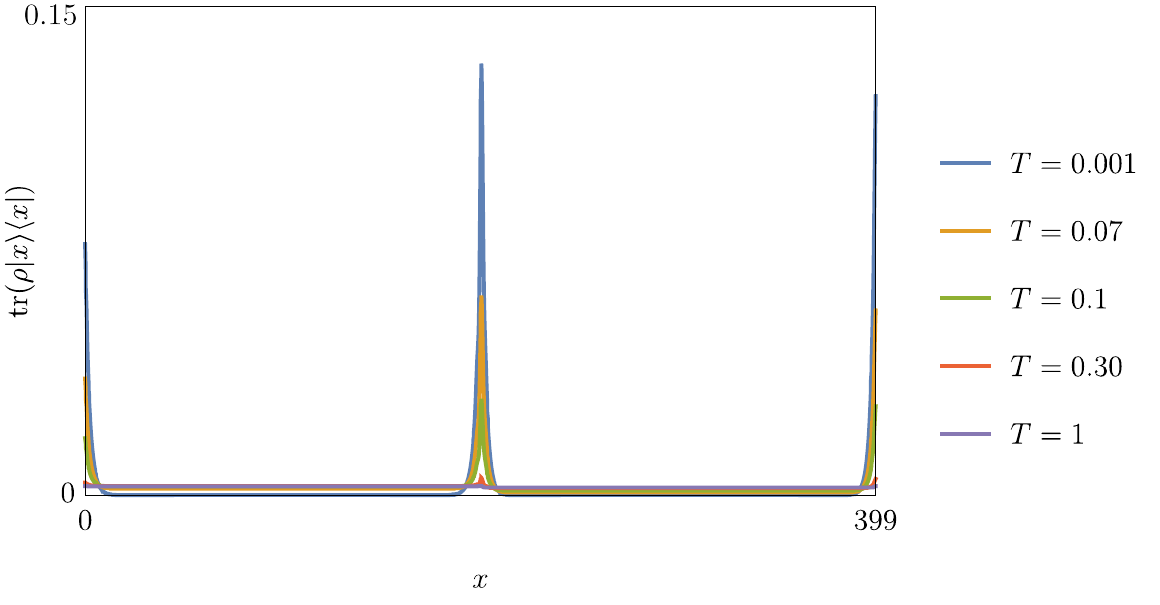}
\caption{
Position probability distribution of the QW as a function of the sites, $\tr (e^{-H/T}\ket{x}\bra{x})/Z$. The Hamiltonian $H$ is obtained by varying $\theta$ along $x$ through a step-like function~\cite{kit:rud:ber:dem:10}. The domain wall is centred in the middle of the line. Periodic boundary conditions are taken, hence the edge state at the boundary.}
\label{fig:edge}
\end{figure}

The behaviour of the edge states, as temperature increases, is shown in Fig.~\ref{fig:edge}:
\\

\subsection*{4. Comment on Figure 1}
  In the fidelity plot for the case of many-body $\varrho^{(1)}$ state, we observe a step-like behaviour at the critical point $(\theta_c,k_c)=(-\frac{\pi}{2},\pi)$. This is due to numerical instabilities occurring close to zero temperature. In the following figure, we plot with higher precision ($\delta\theta=\delta T=\delta k=0.001,$ instead of 0.01) a small region $T\in [0.003,0.3]$ and $\theta\in [-\frac{\pi}{2}-0.2,-\frac{\pi}{2}+0.2]$ around this ``step''. We observe that this behaviour is not present anymore, confirming that it is indeed a numerical instability and it is not associated to a physical phenomenon. Since the higher precision is relevant only around this single point $(\theta_c,k_c)$, and the higher precision calculations are time consuming (it took about two days to obtain this small fraction of the whole plot), we kept the lower precision plots in Fig. 1.
  
\begin{center}
\includegraphics[scale=0.6]{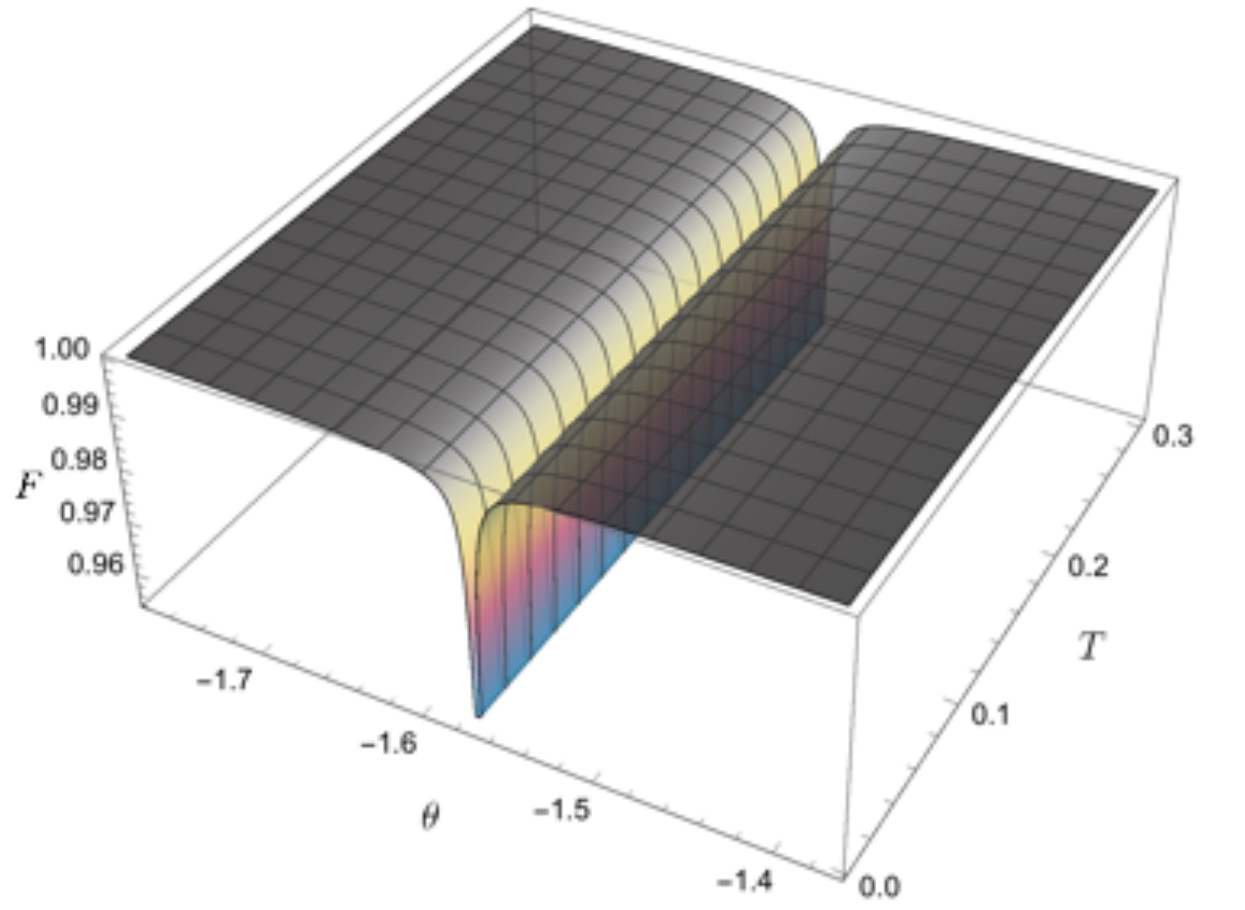}
\end{center}

\bibliographystyle{unsrt}
\bibliography{QW_paper}
\end{document}